\documentclass[preprint2]{aastex}

\usepackage[varg]{txfonts}
\usepackage{graphicx}
\usepackage{natbib}
\usepackage{morefloats}
\usepackage[utf8]{inputenc}
\usepackage{ textcomp }
\usepackage{ gensymb }  
\usepackage{hyperref}   
\usepackage{float}

\begin{document}

   \title{Evidence for evaporation-incomplete condensation cycles in warm solar coronal loops}

   \author{C.~Froment\altaffilmark{1}, F.~Auch\`ere\altaffilmark{1}, K.~Bocchialini\altaffilmark{1}, E.~Buchlin\altaffilmark{1}, C.~Guennou\altaffilmark{2,3} and J.~Solomon\altaffilmark{1} }     
   \affil{\altaffilmark{1}Institut d'Astrophysique Spatiale, B\^atiment 121, CNRS/Universit\'e Paris-Sud, F-91405 Orsay cedex, France} 
   \affil{\altaffilmark{2}Instituto de Astrof\'isica de Canarias,  C/ V\'ia L\'actea, s/n, E-38205, La Laguna, Tenerife, Spain}
   \affil{\altaffilmark{3}Departamento de Astrof\'isica, Universidad de La Laguna, E-38206 La Laguna, Tenerife, Spain}
   \email{clara.froment@ias.u-psud.fr} 
   
   \affil{Accepted for publication in the Astrophysical Journal 2015 April 16}
   
   
   \begin{abstract}Quasi-constant heating at the footpoints of loops leads to evaporation and condensation cycles of the plasma: thermal non-equilibrium (TNE). This phenomenon is believed to play a role in the formation of prominences and coronal rain. However, it is often discarded to be involved in the heating of warm loops as the models do not reproduce observations. Recent simulations have shown that these inconsistencies with observations may be due to oversimplifications of the geometries of the models.
In addition, our recent observations reveal that long-period intensity pulsations (several hours) are common in solar coronal loops. These periods are consistent with those expected from TNE. The aim of this paper is to derive characteristic physical properties of the plasma for some of these events to test the potential role of TNE in loop heating.
We analyzed three events in detail using the six EUV coronal channels of SDO/AIA. We performed both a Differential Emission Measure (DEM) and a time-lag analysis, including a new method to isolate the relevant signal from the foreground and background emission.
For the three events, the DEM undergoes long-period pulsations, which is a signature of periodic heating even though the loops are captured in their cooling phase, as is the bulk of the active regions.
We link long-period intensity pulsations to new signatures of loop heating with strong evidence for evaporation and condensation cycles. We thus witness simultaneously widespread cooling and TNE. Finally, we discuss the implications of our new observations for both static and impulsive heating models.
   \end{abstract}
    
                      
   \maketitle
 
\newcommand*\ebcomment[1]{\textbf{[#1]}}
   
   \section{Introduction}   

\paragraph{}The heating mechanism(s) of coronal loops that generate million-degree plasma and maintain them at such temperatures remain unknown. Some of the main challenges is to match observational results with model predictions and to find signatures specific to some heating process.
Among heating models, scenarios of two different kinds can be differentiated: the impulsive heating scenarios and the quasi-continuous ones.

Small-scale impulsive heating scenarios are known under the term of "nanoflares". A nanoflare consists in a relatively small energy impulsive event occurring by reconnection of loop strands \citep[e.g][]{cargill_1997,patsourakos_2006,klimchuk_2006}. If many of these stochastic events happen close in time in the loops, we have what is called a nanoflare storm. The nanoflare term was first proposed by \citet{parker_1988}, and, since then, signatures of such events are investigated. The heating by nanoflares can produce million-degree type loops \citep{guarrasi_2010} as well as very high temperature loops ($T > 5$~MK) \citep{cargill_1994,klimchuk_2006}, which is one of the main strong points of heating by nanoflares.

The other class of scenarios involve continuous or quasi-continuous heating concentrated at the footpoints of the loops. Numerical simulations show that in that case, the loops can undergo cycles of evaporation and condensation, a phenomenon called thermal non-equilibirum.
Thermal non-equilibirum processes are known to play an important role for prominences \citep[]{antiochos_1991,karpen_2006} and coronal rain \citep{muller_2005}. Coronal rain is considered to be the consequence of "catastrophic cooling" events, as seen in numerical simulations of thermal non-equilibrium \citep{muller_2003,muller_2004}. \citet{antolin_2010} connect coronal rain to coronal heating mechanisms by means of a comparison between observations and MHD simulations of loops heated at the footpoints. This heating produces condensation-evaporation cycles that, in case of a strong plasma condensation at the loop apex, form "blobs" of cold plasma ($T~\sim 0.1$~MK).
Motion of these blobs were observed by \citet{de_groof_2004} with the Extreme Ultraviolet Imaging Telescope \citep[EIT;][]{eit1995} on board the Solar and Heliospheric Observatory \citep[SoHO;][]{soho1995} and \citet{schrijver_2001} with the Transition Region And Coronal Explorer \citep[TRACE;][]{handy_1999}.

\citet{mok_2008} consider that thermal non-equilibrium could be involved also in the formation of coronal loops. In their simulations, a heating concentrated at the footpoints leads to condensation-evaporation cycles. The temperature oscillates in time with a period of about 16 hours. However, the role of thermal non-equilibrium in heating of warm coronal loops has often been discarded, as simulations are not able to reproduce all the observational constrains \citep{klimchuk_2010}.

Nevertheless, oversimplifications of the model geometry could explain these inconsistencies with observations. \citet{lionello_2013} show an analysis of synthetic images of an active region, obtained from 3D-hydrodynamic simulations, that seems consistent with observational results. In their paper, the authors analyzed emission from loops in the same way as if they analyzed real images with diagnostics confirmed by \citet{winebarger_2014}. They successfully checked that the loops extracted reveal seven fundamental characteristics of warm loops. In this series of papers, \citet{mikic_2013} explore influences of the geometry of the loops and of the symmetry of heating in 1D models. Some cases lead to "catastrophic cooling" events that they called "complete" condensations. However, with other configurations loops stay stable to temperatures above 1~MK when only "incomplete" condensations are formed.

\paragraph{}Recently, long-period (several hours) intensity pulsations have been found to be very common in active regions, especially in loops \citep{auchere2014}. These pulsations are likely to be new observational signatures of the heating processes in coronal loops.

In this paper we analyze in detail some long-period pulsation events found in loops to investigate the physical mechanisms involved, using data from the Atmospheric Imaging Assembly \citep[AIA;][]{boerner2012sdo,lemen2012sdo} on board the \textit{Solar Dynamics Observatory} \citep[SDO;][] {pesnell2012sdo}. This allows to track long-period intensity pulsation events simultaneously in six coronal extreme ultraviolet (EUV) channels: 94~\AA, 131~\AA, 171~\AA, 193~\AA, 211~\AA, 335~\AA. Compared to the previous analysis in one EIT band \citep{auchere2014}, we can thus now perform physical diagnostics of the plasma. The three cases that we present here show strong evidence for cycles of evaporation and of incomplete condensation.
First we present the analyzed data in Sec.~\ref{sec:data}. Then we move to the physical analysis with a DEM diagnostic in Sec.~\ref{sec:dem} and to giving evidence for widespread cooling in Sec.~\ref{sec:cooling}.

   \section{Analysis of 3 typical loop events}  

     
     \subsection{Data sample}\label{sec:data}
       
	\begin{figure*}
		\centering
                 \includegraphics[width=\linewidth, trim = 0 0 0 1.92cm, clip]{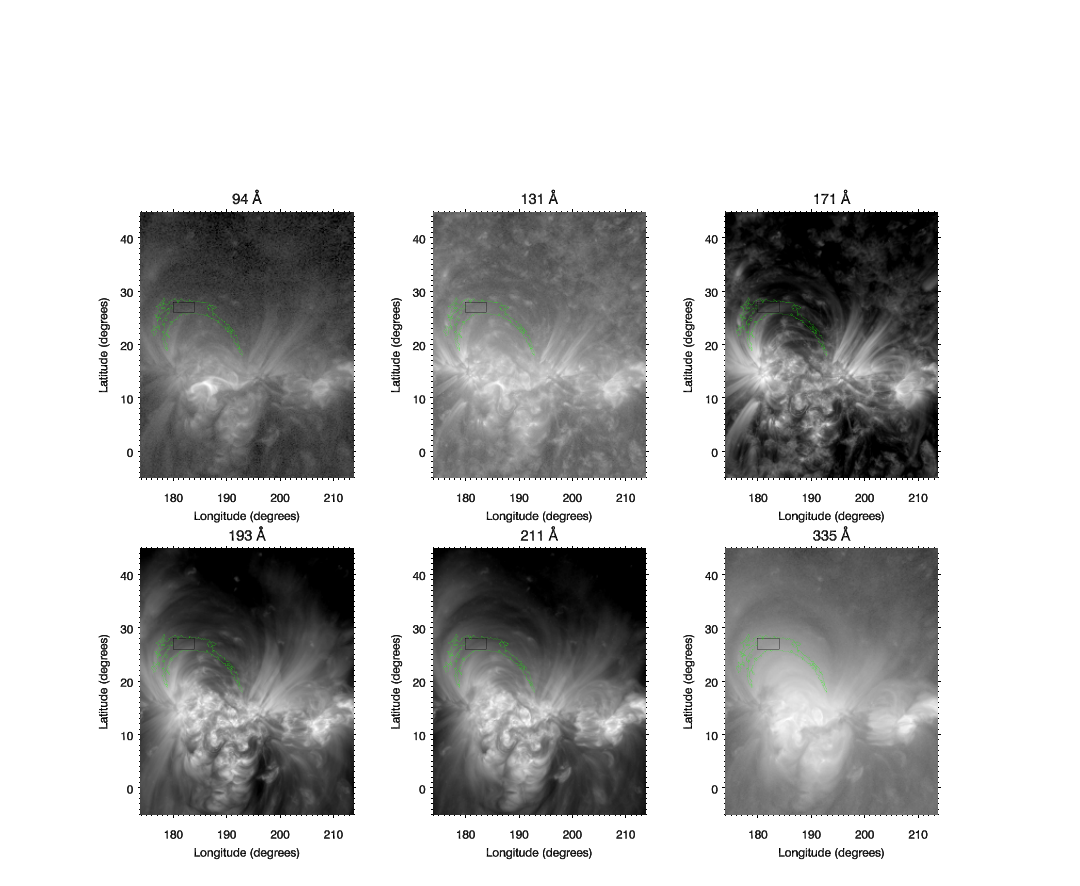}
                 \caption{Individual frames for event~1 for the six coronal channels of SDO/AIA on June 07, 2012 at 13:47 UT. This event is localized in NOAA AR 11499. These images are displayed with a logarithmic scale and in heliographic coordinates. The green contour delimits the area of the pulsations detected at 335~\AA. The black contour delimits the area manually selected for detailed time series analysis.}
                 \label{fig:event_1_images}
	\end{figure*}
        
\paragraph{}In order to detect long-period intensity pulsation events, we use the automatic detection algorithm developed by \citet{auchere2014}. The algorithm treats sequences of about 6 days of data at a cadence of 13 minutes, which represents about 600 images. For each sequence, a region of interest (ROI) is tracked by mapping these images into heliographic coordinates (with a resolution of 0.2\(\degree\) in longitude and latitude) and by compensating for the differential rotation. The data cube is analyzed in the Fourier space (obtained by Fourier transform of the data cube along the time axis) with respect to several criteria, the main one being a detection threshold at 10\(\sigma\) above an estimate of the average local power. Using the database of images of EIT/SoHO in the 195~\AA~passband, from January 1997 to July 2010, \citet{auchere2014} reported 917 events of long-period intensity pulsations. About half of these events have been detected in active regions and \(50\%\) of these have been visually associated with loops.

\paragraph{}We analyze here data from AIA using the same detection algorithm. The AIA raw data are read with the routine \texttt{read\_sdo} from the Interactive Data Language SolarSoftware library. We normalize the intensities by the exposure time and so they are expressed in DN.pix\(^{-1}\).s\(^{-1}\) (DN: Digital Number). In order to increase the signal-to-noise ratio, we start from \(4 \times 4\) binned images since the detected regions are large (larger than the typical area of bright points). In addition, given the detected periods, we use a cadence of 13 min. The differential rotation is compensated for all the passbands with the rate measured in the 195~\AA~passband of EIT\footnote{We assume that the structures rotate with the same velocity independently of the channel used.}\citep[]{Hortin03, auchere2014}, which corresponds to the average 1 MK corona. For more details about the detection algorithm, the readers should refer to \citet{auchere2014}.
For the Fourier analysis (sections~\ref{sec:data} and \ref{sec:dem}) and all the cross-correlation computations, we resample the time series using a linear interpolation to ensure a regular cadence. We checked that the resampling does not affect the detections of long-period pulsations in the Fourier analysis, as it affects only high frequencies. See \citet[Appendix A: Sources of spurious frequencies]{auchere2014} for details about sources of spurious detections. Since the time series have about 300 images (sequences of three days of data), the \(10\sigma\) detection threshold corresponds to a confidence level of \(99\%\).

\paragraph{}From May 2010 to December 2013 the code detected about 2000 events. As for the EIT based study, 50\(\%\) of them are localized in active regions and about a half of them are visually associated with loops. Since we detect some events that have also been detected with EIT (in May 2010), we can discard definitely instrumental artefacts as the cause for these long-period intensity pulsations. Furthermore, since we do not apply any filter to the data before Fourier analysis, it is also unlikely that these events are an artefact of processing.
Among the detected events, we choose to analyze three events showing a strong signal in loops, in order to investigate the underlying physical processes in detail.
These three events cover a large range of periods. For the first event, the power peaks around 30.7~\(\mu\)Hz, i.e. 9.0 hours, for the second around 49.9~\(\mu\)Hz, i.e. 5.6 hours and for the third around 72.9~\(\mu\)Hz, i.e. 3.8 hours. We can thus explore the possible relationship between frequency and physical properties. The three active regions with pulsating loops are also very different with respect to their size. In addition, the three tracked active regions have a moderate activity with no flares above C-class.
To simplify the reading and because we reach the same conclusions with the three events, we only present event~1 in detail in the core of the text. Events~2 and 3 are presented respectively in Appendix~\ref{sec:event2_figures} and Appendix~\ref{sec:event3_figures} with the same progression. 

	\begin{figure*}
		\centering
                 \includegraphics[width=\linewidth]{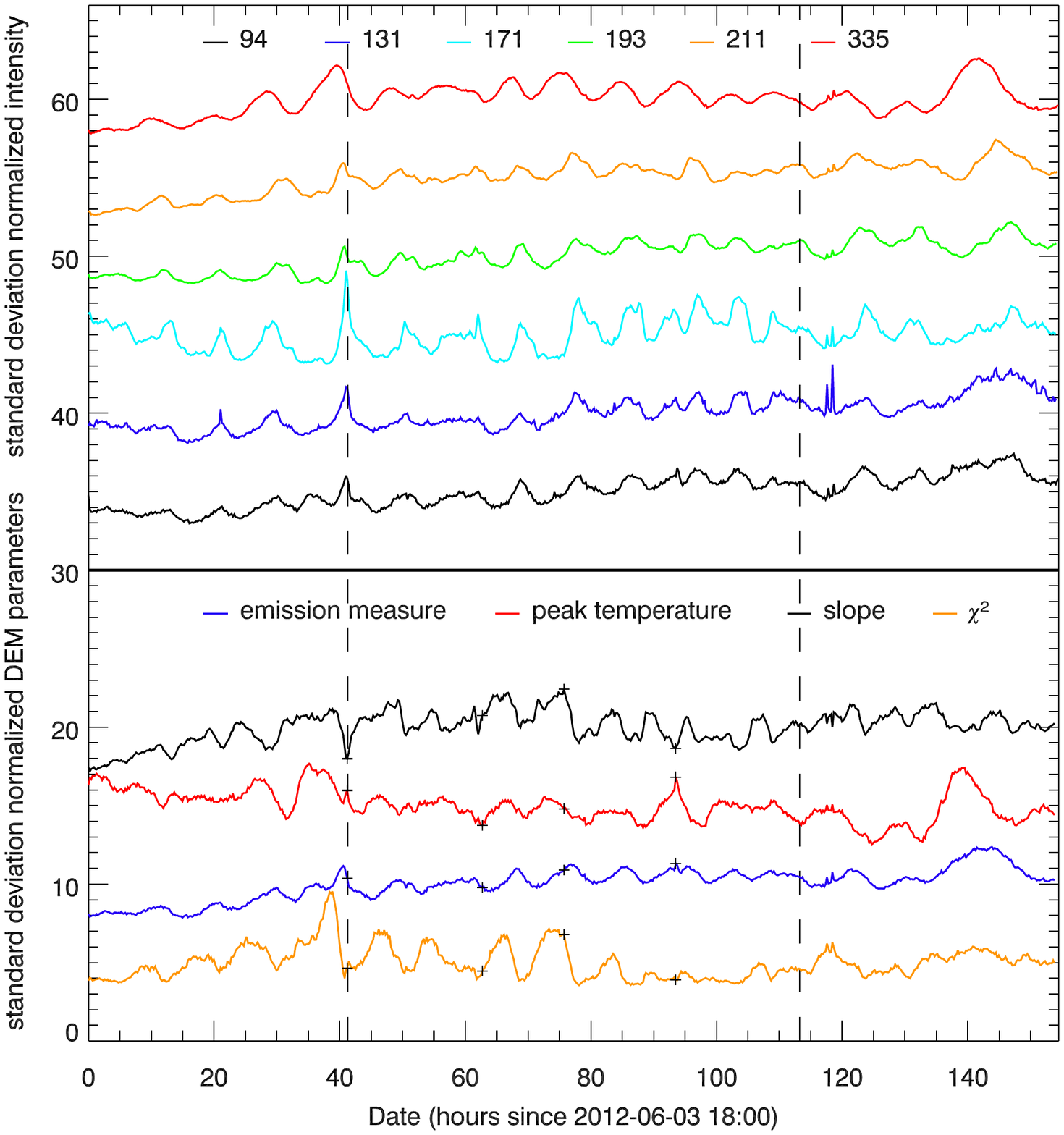}
                 \caption{The six light curves and the evolution of the three free parameters of the DEM model (plus \(\chi^{2}\)) from June 03, 2012 18:00~UT to June 10, 2012 04:29~UT for event~1. We have used mean intensities and mean DEM parameters in the small black contour (Fig.~\ref{fig:event_1_images}). All these curves are normalized to standard deviation and offset by 5.0 along the y-axis. Between DEM curves on the bottom and light curves on the top, there is an offset of 15.0 in the y-axis. We restrict our analysis to the middle of the sequence (marked by the vertical dashed lines) in order to minimize distortion effects (see section~\ref{sec:data} for details). Plus signs indicate instants chosen for the four cases of DEM shape in Fig.~\ref{fig:dem_shapes_event_1}.}
                 \label{fig:light_curves_1}
	\end{figure*}

\paragraph{} Fig.~\ref{fig:event_1_images} represents the heliographic field of view for event~1 on June 07, 2012 at 13:47~UT in the six coronal channels of SDO/AIA. We tracked NOAA AR~1499 during 154 hours (i.e. more than 6 days), from June 03, 2012 18:00 UT to June 10, 2012 04:29 UT. In order to minimize distortion effects due to the heliographic mapping transformation\footnote{Distortion of the structures is more important close to the limb and if the structures are tracked for a long time.}, we kept only three central days in this sequence of six days, from June 05, 2012 11:14~UT to June 08, 2012 11:16~UT. For the rest of the data analysis we will use this short sequence, except if otherwise stated. The green contour in Fig.~\ref{fig:event_1_images} delimits the area of pulsations detected at 335~\AA, which is the passband with the strongest signal (see. Fig.~\ref{fig:event_1_power_maps}). This contour is centered on 182.9\(\degree\) of longitude and 25.3\(\degree\) of latitude and has an area of 45.5 heliographic square degrees (corresponding to 5271 Mm\(^{2}\) on the sphere). This contour fits the shape of large loops visible in images. We manually selected a smaller contour in black to delimit the area used for detailed time series analysis. This contour is included in the contour detected at 335~\AA~and the signal is strong for most of the passbands: more than \(7\sigma\) of normalized power, averaged over this contour for five passbands (see Fig.~\ref{fig:event_1_power_maps}). We choose this contour so that the phases of the periodic components (see Fig.~\ref{fig:event_1_phase_maps}) are similar in order to maximize the signal-to-noise ratio when averaging over severals heliographic pixels.
        
	\begin{figure*}
		\centering
                 \includegraphics[width=\linewidth]{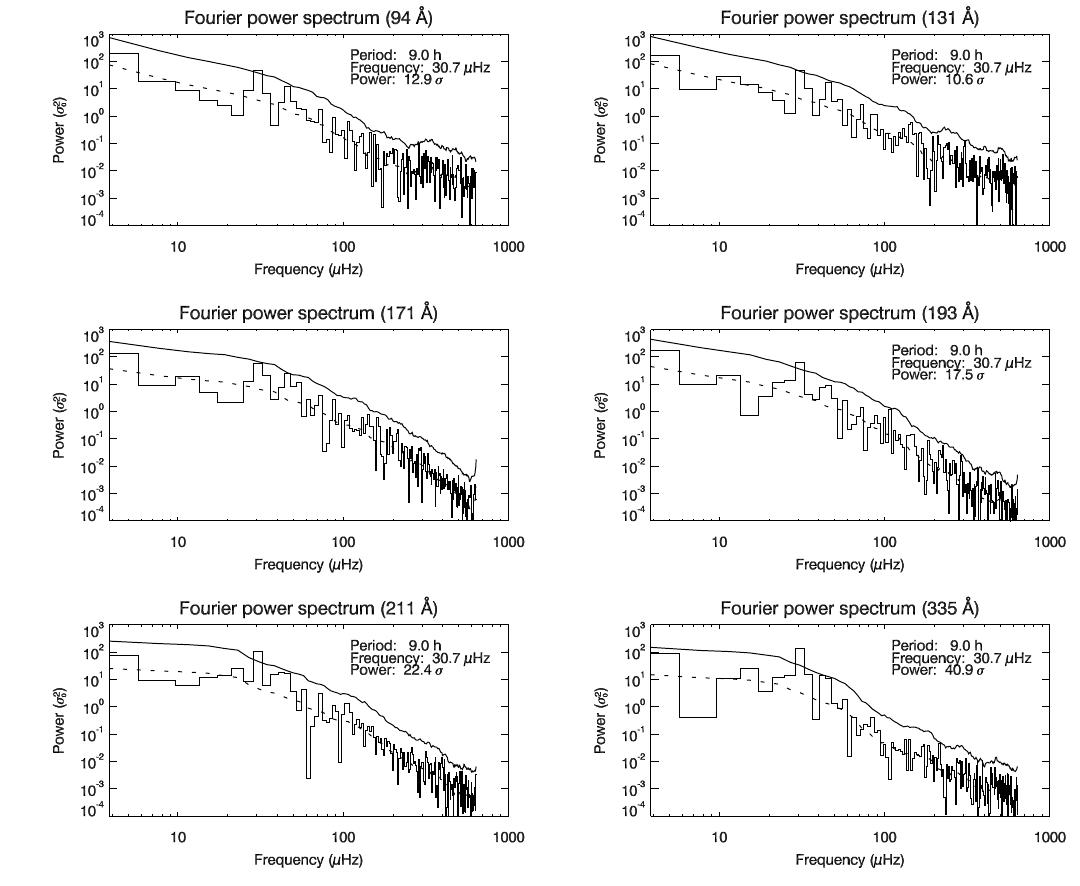}
                 \caption{Fourier power spectra for the six AIA coronal channels. These spectra are computed for the light curves between the two vertical dashed lines in Fig.~\ref{fig:light_curves_1}. The dashed line is the estimate of the average local power. The solid line is the 10\(\sigma\) detection threshold. If a peak of power is above the detection threshold, we display the corresponding period of pulsations and normalized power at the top right of each spectrum.}
                 \label{fig:event_1_spectrum}
	\end{figure*}
        
We choose to concentrate on time series of the average parameters in this black contour, because isolating pulsating loops and subtracting a background (albeit crucial for most studies of coronal loops) is neither practical nor meaningful for the present analysis.
Actually, our time series are much longer than the typical life time of loops: 1000\---5000 s \citep[]{winebarger_et_al_2003,winebarger_warren_2005,ugarte-urra_et_al_2009}, therefore it is not possible to isolate and to track loops over such long times. 
In Fig.~\ref{fig:light_curves_1}, we display the six light curves averaged over this black contoured area: 94~\AA~in black, 131~\AA~in blue, 171~\AA~in cyan, 193~\AA~in green, 211~\AA~in orange and 335~\AA~in red. These light curves are normalized to standard deviation (we subtract the mean curve and divide by the standard deviation). The vertical dashed lines delimit the short sequence (i.e. 3 days) used to compute the Fourier spectra.

Fig.~\ref{fig:event_1_spectrum} represents the corresponding Fourier power spectra, normalized to the variance \(\sigma_{0}^{2}\) of these light curves \citep{torrence98}. The dashed line is the estimate of the average local power. The solid line is the 10\(\sigma\) detection threshold\footnote{\(\sigma\) is the averaged power as a function of frequency and is different from \(\sigma_{0}^{2}\), which is the variance of the time series. In the case of white noise, we would have \(\sigma = \sigma_{0}^{2}\) }. We consider only peaks of power between 18~\(\mu\)Hz and 110~\(\mu\)Hz (between 15.4 and 2.5 hours). In the 94~\AA, 131~\AA, 193~\AA, 211~\AA, and 335~\AA~channels there is a significant peak of power at 30.7~\(\mu\)Hz (9.0 hours). The normalized power at these peaks are respectively 12.9\(\sigma\), 10.6\(\sigma\), 17.5\(\sigma\), 22.4\(\sigma\), and 40.9\(\sigma\). 
In the 171~\AA~passband there is no peak of normalized power above 10\(\sigma\). However, there is a 8.9\(\sigma\) peak at the frequency detected in the other bands (30.7~\(\mu\)Hz), plus a second 9.3\(\sigma\) peak at 46.1~\(\mu\)Hz (6.0 hours). This small peak at 6.0 hours exists also in the 94, 131, 211, and the 335 channels with respectively 7.4\(\sigma\), 8.3\(\sigma\), 8.0\(\sigma\), and 8.9\(\sigma\) of normalized power. All these peaks correspond to confidence levels greater than \(90\%\). At 211~\AA~and 335~\AA, we can also notice small peaks at 42.2~\(\mu\)Hz with respectively 7.4\(\sigma\) and 9.2\(\sigma\) of normalized power.

Fig.~\ref{fig:event_1_power_maps} represents the maps of normalized power in heliographic coordinates for the six passbands at the frequency of detection (30.7~\(\mu\)Hz, i.e. 9.0 hours), with the same field of view in Fig.~\ref{fig:event_1_images}. The normalized power is displayed in logarithmic scale, with a saturation at 20\(\sigma\). As for Fig.~\ref{fig:event_1_images}, we displayed the contour detected at 335~\AA~in green and the contour that we use for analysis in black. We can notice that the power is well localized in loops for most of the passbands. In the black contour, the normalized power is higher or equal to 7\(\sigma\) on average for all the passbands, except for 131~\AA~ the power is the weakest (5\(\sigma\) in average). This value rises to 19\(\sigma\) at 335~\AA.
        
	\begin{figure*}
		\centering
                 \includegraphics[width=\linewidth, trim = 0 0 0 1.92cm, clip]{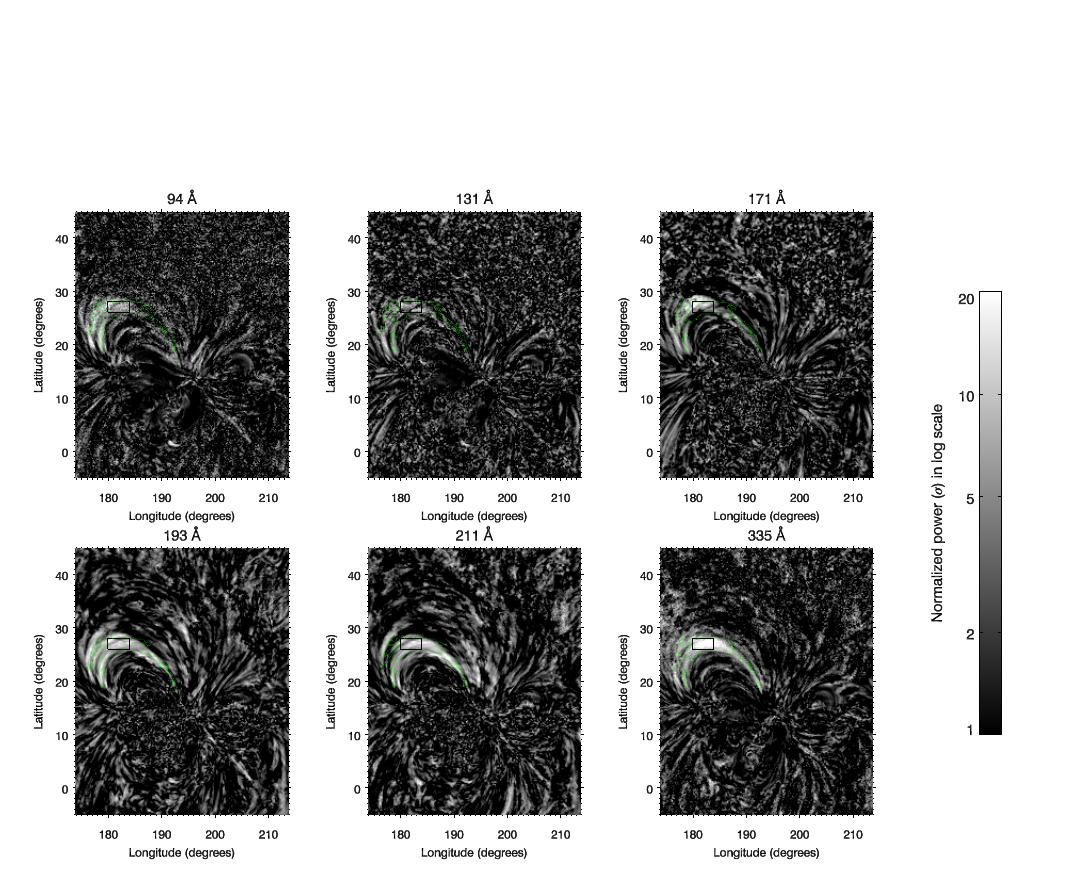}
                 \caption{Maps of normalized power at 30.7~\(\mu\)Hz (i.e. 9.0 hours) for event~1. These maps are in heliographic coordinates for the six EUV passbands of AIA: 94~\AA, 131~\AA, 171~\AA, 193~\AA, 211~\AA, 335~\AA. The normalized power is in logarithmic scale, from 1\(\sigma\) to 20\(\sigma\). The black contour, included in the detected contour in green, is the one used for detailed time series analysis.}
                 \label{fig:event_1_power_maps}
	\end{figure*}
                
     \subsection{Differential emission measure (DEM) analysis}\label{sec:dem}
        \subsubsection{Method} 
\paragraph{}We compute the differential emission measure (DEM) using the method developed in \citet{guennou2012_1,guennou2012_2}. To fit the DEM, we choose the active region model developed by \citet{guennou2013} with the six coronal EUV passbands of AIA. This parametric model is designed to represent the shape of the DEM usually observed in active regions \citep[]{warren2011,winebarger2011}. Four parameters determine the shape (as in Fig.~\ref{fig:dem_shapes_event_1}) of this DEM model: 
\begin{itemize}
        \item the slope \(\alpha\) of the low temperature wing represented by a power law 
        \item the peak temperature of the DEM
        \item the total emission measure (the integral of the DEM over the electron temperature \(T_{e}\))
        \item the width \(\sigma\) of the high temperature Gaussian wing. We choose to set this parameter to a fixed value of \(0.1 \log T\) as the high temperature wing of the DEM is poorly constrained by AIA data.
\end{itemize}

We compute the DEM using the six channels of AIA. For this, we grouped images from the six passbands into sextuplets allowing for a maximum non simultaneity of \(\pm 2\) minutes.
 
        \subsubsection{Results}

	\begin{figure*}
		\centering
		\includegraphics[width=\linewidth]{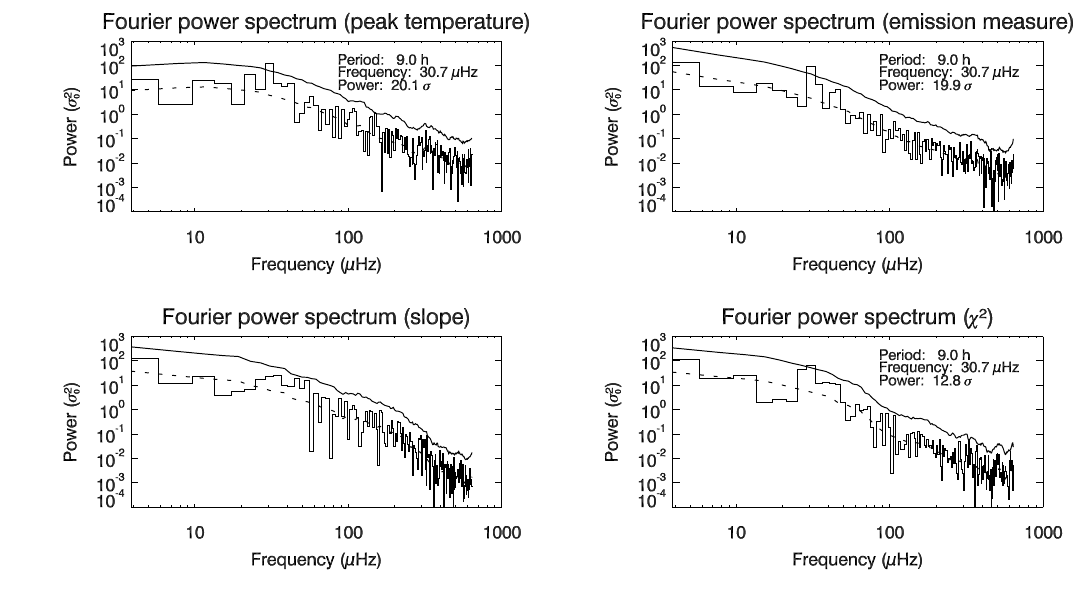}
		\caption{Fourier power spectra for the three parameters of the DEM model for event 1: the peak temperature, the emission measure and the slope. We represent also the Fourier power spectrum for \(\chi^{2}\). These spectra are computed for light curves between the two vertical dashed lines in Fig.~\ref{fig:light_curves_1}. The dashed line is the estimate of the average local power. The solid line is the 10\(\sigma\) detection threshold. If a peak of power is above the detection threshold, we display the corresponding period of pulsations and normalized power at the top right of each spectrum.}
                 \label{fig:dem_event_1_spectrum}
	\end{figure*}

\paragraph{} For event~1, the evolution of the slope, peak temperature, emission measure, and \(\chi^{2}\) residuals (a measure of the adequacy of the chosen model to represent the real DEM), averaged over the black contour introduced in Fig.~\ref{fig:event_1_images}, is presented in Fig.~\ref{fig:light_curves_1}.
Fig.~\ref{fig:dem_event_1_spectrum} represents the power spectra of these time series. The peak temperature, the emission measure and \(\chi^{2}\) present pulsations at the same frequency as the intensity. Actually there are power peaks at 30.7~\(\mu\)Hz (9.0 hours) with respectively 20.1\(\sigma\), 19.9\(\sigma\) and 12.8\(\sigma\) for these two DEM model parameters and for \(\chi^{2}\). For the slope there is no significant power peak but a small peak (7.4\(\sigma\)) at 46.1~\(\mu\)Hz, the frequency of the secondary peaks found in the power spectra of the light curves (see Fig.~\ref{fig:event_1_spectrum}).
Even if we found pulsations in \(\chi^{2}\), the absolute values of \(\chi^{2}\) varie between 0.2 and 1.5, which indicates a generally good fit to the DEM. 
        
	\begin{figure}[h!]
		\centering
                 \resizebox{\hsize}{!}{\includegraphics{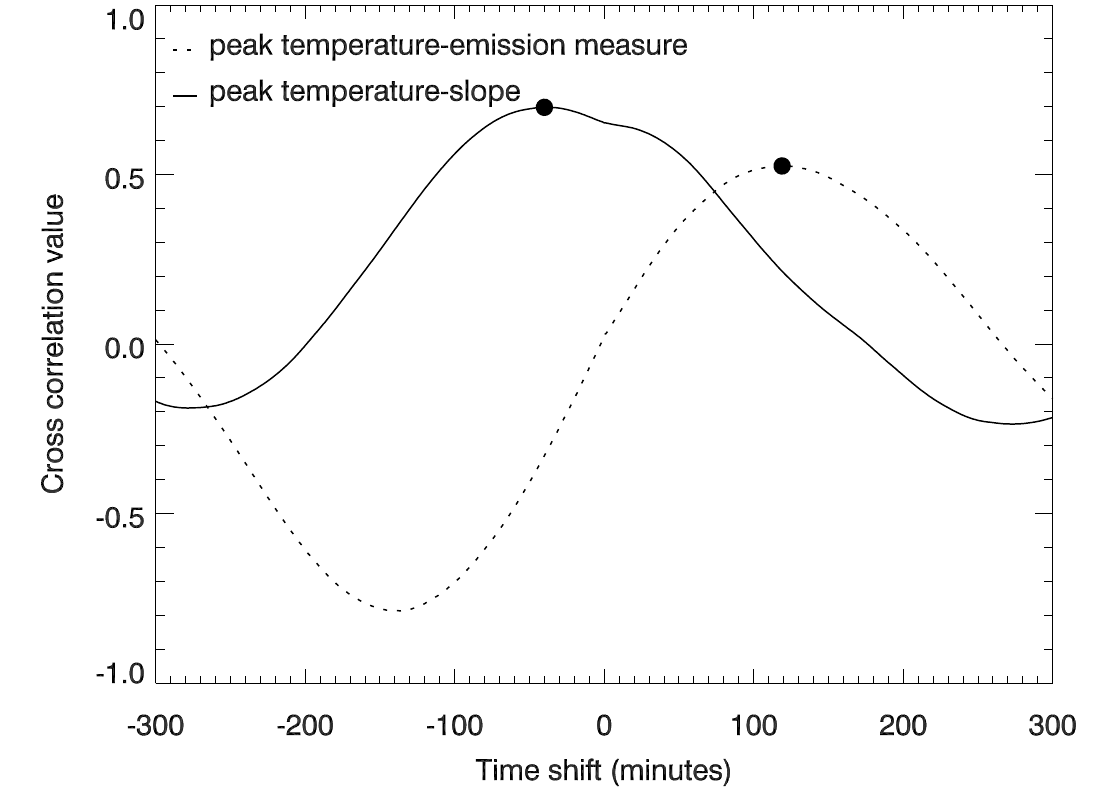}}
                 \caption{Cross-correlation values for peak temperature/slope in solid line and peak temperature/emission measure in dotted line. We explored time shifts from -300 minutes to 300 minutes. Time lag for each pair of DEM parameters is indicated by a black dot. We used the averaged curves at 1 min of cadence averaged over the small black contour presented in Fig.~\ref{fig:event_1_images} between June 06, 2012 08:42~UT (63 h after the beginning of the long sequence) and June 07, 2012 12:00~UT (90 h after the beginning of the long sequence).}
		\label{fig:correlations_dem_event_1}
	\end{figure}

\paragraph{}Although there is no significant power at the frequency detected in the intensity time series for the DEM slope, we found a correlation between the slope and both the emission measure and the temperature, as seen in Fig.~\ref{fig:correlations_dem_event_1}. The cross-correlation between the peak temperature and the emission measure is in dotted line and that between the peak temperature and the slope is in solid line with a time shift between \(-300\)~min and \(+300\)~min. These curves are computed over a duration of about 28 hours, between 62.7 hours after the beginning of the sequence (i.e. on June 06, 2012 08:42~UT) and 90 hours after the beginning of the sequence (i.e. on June 07, 2012 12:00~UT). There is a time lag of -40 minutes for a peak cross-correlation value of 0.7. This means that variations of the slope precede variations of the peak temperature. Between the peak temperature and the emission measure, there is also a correlation. The peak temperature is in advance of the emission measure with a time lag of 119 minutes for a peak cross-correlation value of 0.5. 

	\begin{figure}[h!]
		\centering
                 \resizebox{\hsize}{!}{\includegraphics{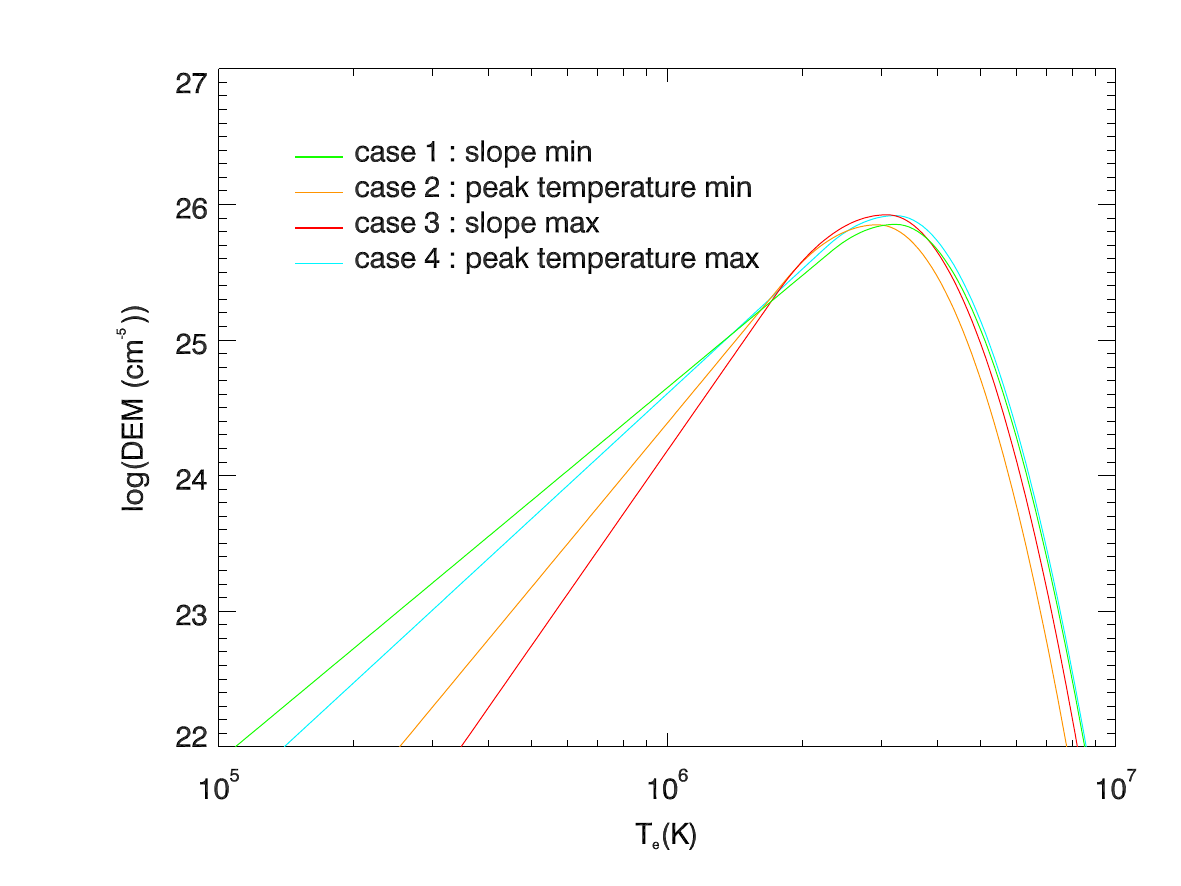}}
                 \caption{Four extreme cases of the shape of the DEM model (see Fig.~\ref{fig:light_curves_1}). In green, the shape of the DEM on June 05, 2012 11:14 UT with the minimum slope (case 1) and in red on June 06, 2012 21:42~UT with the maximum slope (case 3). In orange, the shape of the DEM on June 06, 2012 08:42~UT with the minimum peak temperature (case 2) and in blue on June 07, 2012 15:30~UT with the maximum peak temperature (case 4). The four dates corresponding to these cases are represented by plus signs in Fig.~\ref{fig:light_curves_1}. }
              	  \label{fig:dem_shapes_event_1}
	\end{figure}

\paragraph{}If we look at the variations of the absolute values of the DEM parameters (instead of normalizing them by their standard deviation, as in Fig.~\ref{fig:light_curves_1}), we can notice that the amplitude of the variations of the peak temperature is relatively small compared to the variations of the emission measure and of the slope: the peak temperature varies from 2.8 MK to 3.1 MK (a relative amplitude of 11\(\%\)), while there are variations from \(2.03 \times 10^{27}\) cm\(^{-5}\) to \(2.82 \times 10^{27}\) cm\(^{-5}\) for the emission measure (relative amplitude of 38\(\%\)) and from 2.7 to 4.7 for the slope (relative amplitude of 71\(\%\)). We can note that larger temperature variations are found for event~2 and event~3 (see Appendix~\ref{sec:event2_figures} and \ref{sec:event3_figures}).

Fig.~\ref{fig:dem_shapes_event_1} represents four extreme cases of the shape of the DEM during the sequence. Instants chosen for the four cases are pointed out sequentially by black plus signs in Fig.~\ref{fig:light_curves_1}. This figure illustrates that, of the three parameters of the DEM model, the slope undergoes the largest relative variations. In case 1 (in green), the DEM model is the most multithermal, as the slope reaches a minimum,  \(\alpha = 2.7\) (on June 05, 2012 11:14 UT), the total emission measure is \(2.49 \times 10^{27}\)~cm\(^{-5}\) and the peak temperature is at 3.0~MK.
In orange (case 2), we trace the shape of the DEM model when the peak temperature reaches a minimum at 2.8~MK. That happens on June 06, 2012~08:42~UT when the slope ${\alpha = 4.0}$ and the total emission measure is \(2.29 \times 10^{27}\)~cm\(^{-5}\). In red (case 3), on June 06, 2012 21:42~UT, the slope is the steepest with \(\alpha = 4.7\). The peak temperature is at 2.9~MK and the total emission measure is \(2.67 \times 10^{27}\)~cm\(^{-5}\). For the case 4 (in blue), the peak temperature reaches a maximum at 3.1~MK. The slope is weaker with \(\alpha = 3.0\) and the total emission measure is \(2.82 \times 10^{27}\)~cm\(^{-5}\). This last case occurs on June 07, 2012 15:30~UT. 
                
\paragraph{}These results show that the DEM changes from a multithermal distribution to a more isothermal distribution and that at least the peak temperature and emission measure vary periodically. In addition, even if the normalized power found in the DEM slope Fourier spectrum is under the detection threshold, there is a clear correlation between slope and peak temperature, which is another indication that the slope also undergoes periodic variations. Therefore, we conclude from the DEM and correlation analysis that there are periodical changes in the thermal structure of the loops where we have detected long-period intensity pulsations. These periodical modifications of the thermal structure must be then a consequence of heating in loops, given that a cooling phase is necessarily preceded by a heating phase. This indicates that these long-period intensity pulsations are linked to loop heating.

     \subsection{Evidence for widespread cooling}\label{sec:cooling}   
        \subsubsection{Method}
\paragraph{}We have seen that the thermal structure of the pulsating loops undergo periodical changes which links these phenomena to loop heating. However when the thermal structure of these loops evolves from a multithermal case to a more isothermal case, we do not observe heating. As was shown by several authors, EUV loops are generally seen in their cooling phase \citep[]{winebarger_warren_2005,warren_et_al_2002,winebarger_et_al_2003,ugarte-urra_et_al_2006,ugarte-urra_et_al_2009,mulu-moore_et_al_2011}. In order to show that this is the case for a large fraction of the active region in which pulsations are detected, we use the method developed in \citet[hereafter V\&K12]{viall&klimchuk2012}. V\&K12 present an analysis of an active region and show that the plasma is mainly in state of cooling, as the EUV intensity peaks first in the hotter passbands and then in the cooler passbands. By computing the time lag between several pairs of channels, V\&K12 obtain the temporal succession of channels and see that the plasma is cooling down. By using 2 hour and 12 hour time series, V\&K12 derived time lags ranging from a couple of minutes to 1.5 hours. 
Since the lines dominating the passbands are formed by collisions, variations of the emission measure produce simultaneous variations in the six coronal passbands. On the contrary temperature variations are expected to be gradually reflected in the passbands according to the ordering of peak temperature response of the channels: 335~\AA~(2.5 MK), 211~\AA~(2 MK), 193~\AA~(1.5 MK), 94~\AA~(1 MK and 7 MK), 171~\AA~(0.8 MK) and 131~\AA~(0.5 MK) \citep[AIA temperature response functions;][]{boerner2012sdo,lemen2012sdo}. 

        \subsubsection{Results}
                
	\begin{figure}[h!]
		\centering
                 \resizebox{\hsize}{!}{\includegraphics{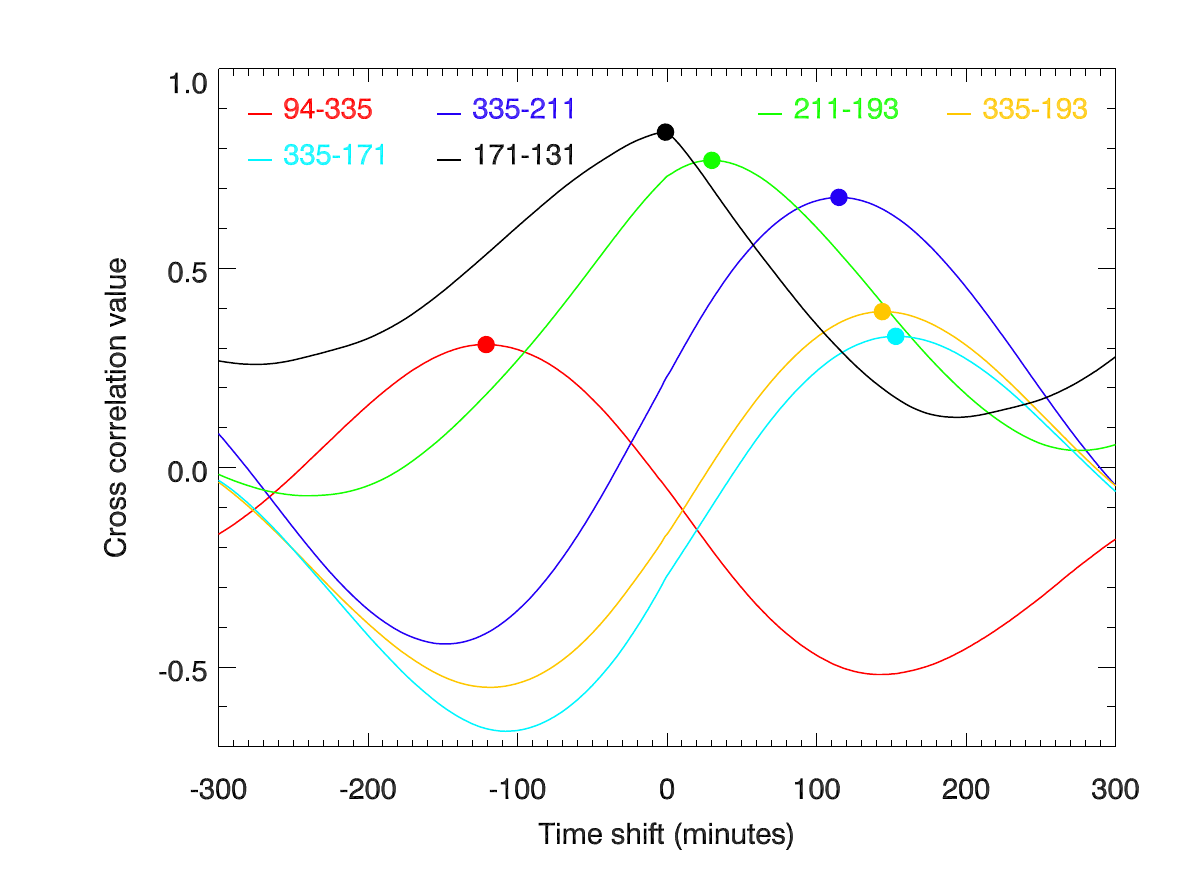}}
                 \caption{Cross-correlation values for six pairs of AIA channels (event~1): 94-335 (red), 335-211 (blue), 211-193 (green), 335-193 (orange), 335-171 (cyan) and 171-131(black). We used light curves at 1 min of cadence averaged over the small black contour presented in Fig.~\ref{fig:event_1_images}. We explored time shifts from -300 minutes to 300 minutes. The time lag for each pair of channel is indicated by a colored dot.}
                 \label{fig:event_1_cooling_curve}       
	\end{figure}

\paragraph{}Our results in Fig.~\ref{fig:event_1_cooling_curve} are to be compared with Fig.~4 of V\&K12. As detailed below, we observe the same behavior as them, meaning that the plasma in the small black contoured area is cooling.

Fig.~\ref{fig:event_1_cooling_curve} represents the cross-correlation values for six pairs of AIA channels: 94-335, 335-211, 211-193, 335-193, 335-171, and 171-131. This calculation is made with the averaged light curves in the small black contour seen in Fig.~\ref{fig:event_1_power_maps}. Here we choose a time cadence of 1 min, as we expect time lags under the 13 min cadence used previously. We explore time shifts from -300 minutes to 300 minutes (i.e. 5 hours). Time lags are given by the peak cross-correlation value for each pair of channels. 

For the 94-335 pair, we obtain a time lag of -121 minutes. This negative time lag indicates that the intensity peaks in the 335 channel before the 94 channel. The temperature response of the 94 channel has two peaks, one cooler than the 335 peak temperature and one hotter.  Since 94 peaks after 335, we conclude that the plasma does not reach the temperature of the hot peak of the 94 band. The 193 channel follows the 211 channel with a time lag of 30 min. For the pairs 335-211, 335-193, and 335-171, time lags are positive and increase gradually, with 115 min, 144 min and 153 min respectively. The 171 and 131 channels show no significant time lag (-1 min) given the time resolution. 

Conclusions are more difficult to draw for the 94-335, 335-193, and 335-171 pairs, which peak cross-correlation values are respectively 0.3, 0.4, and 0.3. Peaks for 335-193 and 335-171 cross-correlations are not well separated and the time lag for 94-335 is smaller in absolute value than the time lag between 335-193 and 335-171. It could be explained because the 94, 171 and 193 channels are the most distant in time compared to the 335 channel: the emission measure is more likely to change during a longer time and the intensities in these passbands are thus more likely to be de-correlated from that in 335. Moreover, we use longer time series that in V\&K12 and thus the passbands are more likely to be de-correlated. Superimposition of structures along the line of sight could also explain this effect since the background contribution may be different in the different passbands. 

	\begin{figure*}
		\centering
                 \includegraphics[width=\linewidth, trim = 0 0 0 1.92cm, clip]{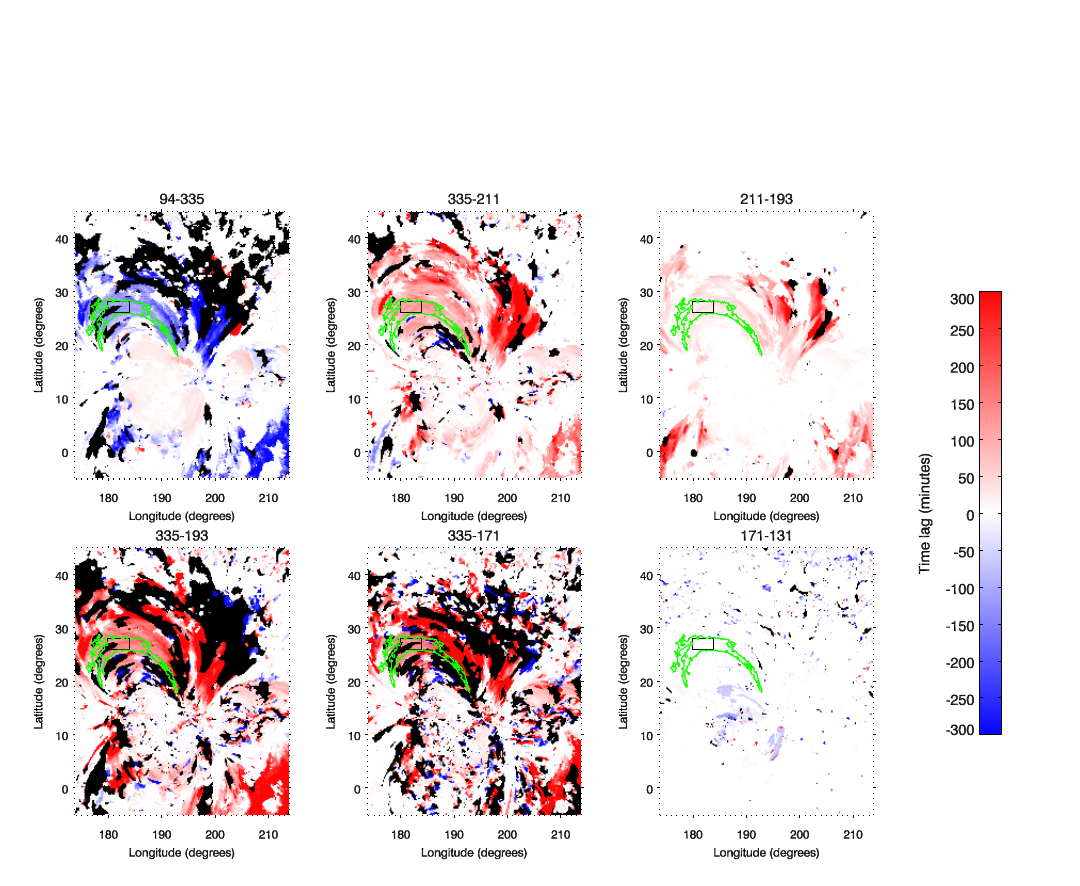}
                 \caption{Time lag maps made with peak cross-correlation values for the same six pairs of AIA channels as in Fig.~\ref{fig:event_1_cooling_curve}: 94-335, 335-211, 211-193, 335-193, 335-171 and 171-131. We explored time shifts from -300 minutes to 300 minutes. Black areas represent peak cross-correlation values under 0.2. The green contour is the detected contour and the black contour is the contour chosen to trace the light curves.}
              	\label{fig:event_1_lag_maps}
	\end{figure*}

\paragraph{}This analysis can be generalized to the whole active region. We build time lag maps like Fig.~5 to 7 of V\&K12. For each pair of channels, the time lag for each pixel is given by the peak cross-correlation value of the intensity time series. Our time lag analysis is fully consistent with the widespread cooling observed by V\&K12. In particular, we observe the same ordering of channels. We can notice that this behavior is found in pulsating areas like in the rest of the active region.

Fig.~\ref{fig:event_1_lag_maps} represents time lag maps for the entire active region. Black areas mark regions where the peak cross-correlation value is under 0.2. In red, the time lag is positive, in blue negative and in white we have a zero time lag. The 335-211, 211-193, 335-193, 335-171 maps are dominated by positive time lags. In the 171-131 map we mainly observe a zero time lag. The 94-335 map reveals two different behaviors. In the core of the active region, the time lag is mainly positive, which indicates that the high temperature component of 94 dominates. On the contrary in the outer loops (including the black contour) the time lag is mainly negative: the low temperature component of 94 dominates. With this method, we do not try to subtract the background and the foreground to isolate the loops. Thus, superimpositions of different structures along the line of sight can occur.

	\begin{figure*}
		\centering
		\includegraphics[width=\linewidth, trim = 0 0 0 1.92cm, clip]{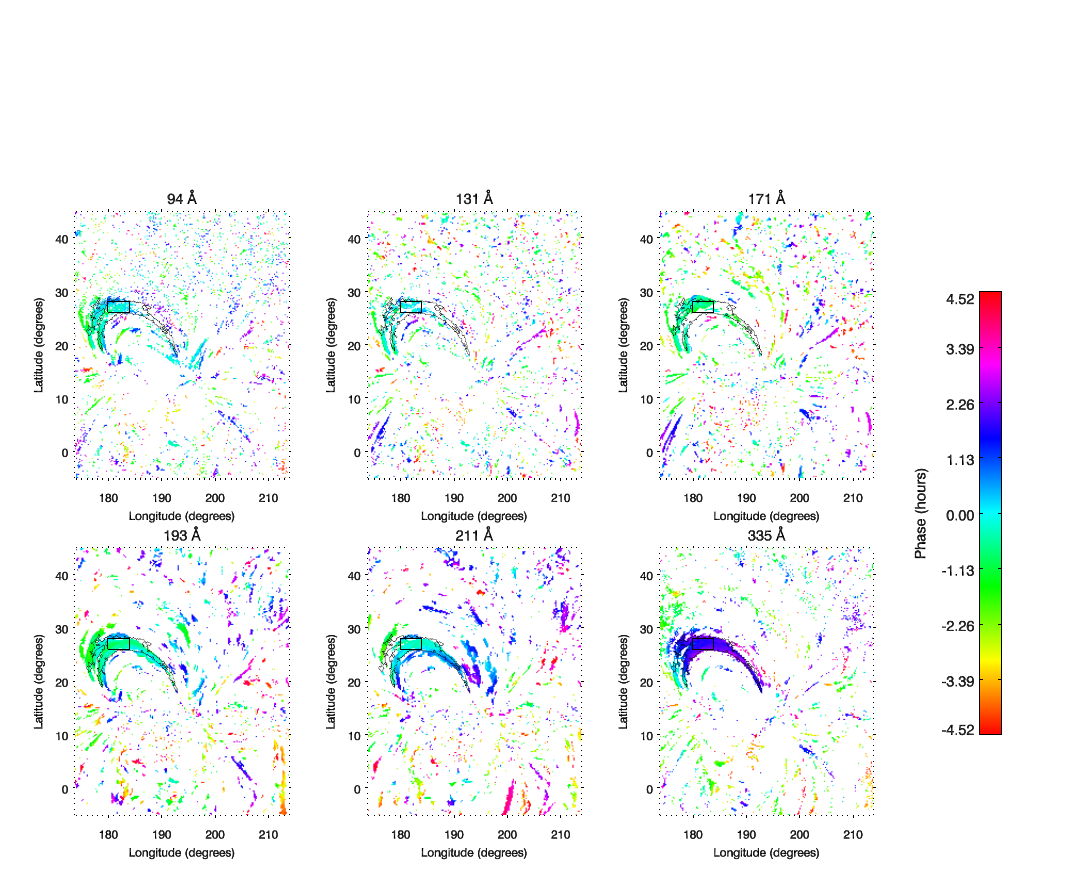}
                 \caption{Phase maps at the frequency of detection (30.7~\(\mu\)Hz, i.e. 9.0 hours) for event~1. The phase is displayed in minutes with a cyclic color table. Here the detected contour is in black to facilitate the reading. Areas with normalized power under 4\(\sigma\) are in white.}
                 \label{fig:event_1_phase_maps}
	\end{figure*}
        
	\begin{figure*}
		\centering
                 \includegraphics[width=\linewidth, trim = 0 0 0 1.92cm, clip]{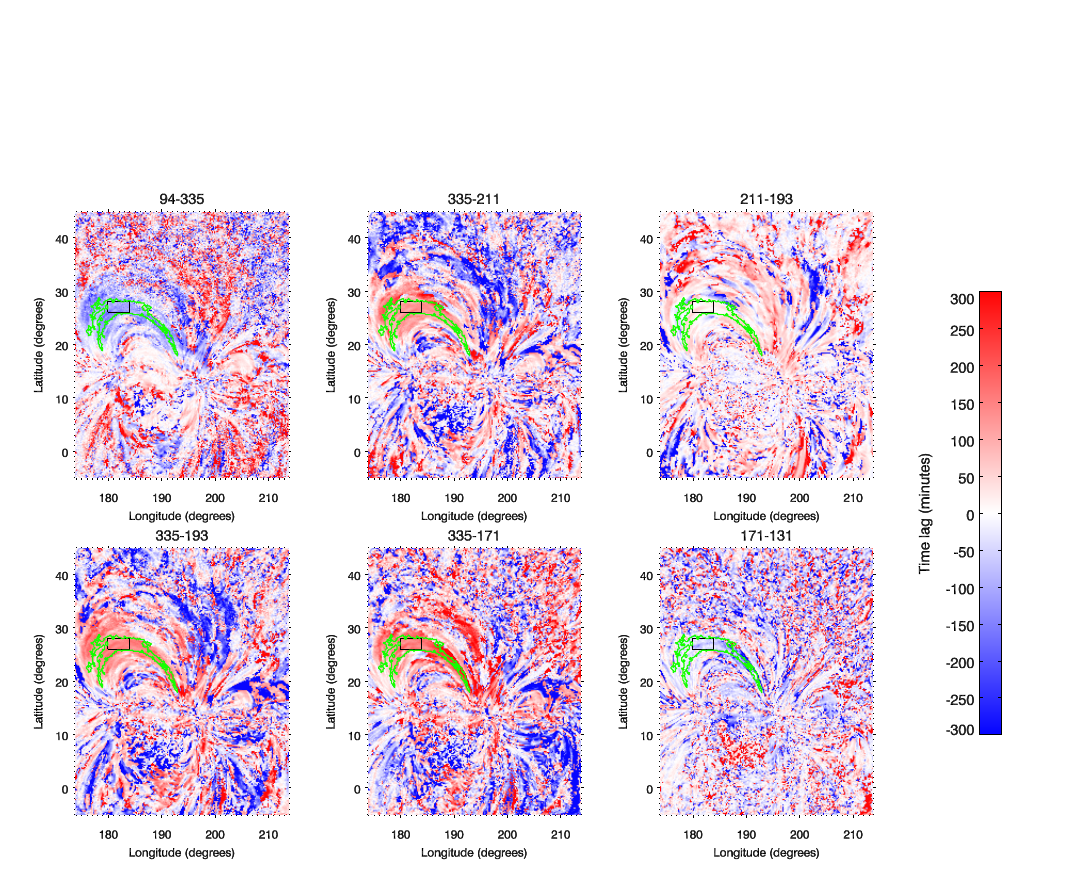}
                 \caption{Time lag maps obtained from differences of phase maps (see Fig.~\ref{fig:event_1_phase_maps}) for the same six pairs of AIA channels as in Fig.~\ref{fig:event_1_cooling_curve}: 94-335, 335-211, 211-193, 335-193, 335-171, and 171-131. We explored time shifts from -300 minutes to 300 minutes. The green contour is the detected contour and the black contour is the contour chosen to trace the light curves.}
                 \label{fig:event_1_phase_maps_lag}
	\end{figure*}
        
\paragraph{}To conclude on the behavior of the pulsating loops with respect to the widespread cooling of the active region, we isolate the pulsating structure in Fourier space: we subtract phase maps at the detected pulsation frequency for pairs of channels, and we convert the phase difference into a time lag. Fig.~\ref{fig:event_1_phase_maps} shows the phase maps. There is no clear evidence for variation of phases along the loops.
Fig.~\ref{fig:event_1_phase_maps_lag} displays the time lag maps obtained with the method of phase maps difference. As can be seen in these maps, the phase difference is only meaningful in areas with detected pulsations; in other regions, the phases are random and so are the time lags.
With this method, the pulsating loops have been isolated, and the background and foreground emission, which are assumed not to be pulsating at the detected frequency, are not taken into account in the calculation of time lags. In addition, we can discern the same pattern of time lags in the pulsating loops as in Fig.~\ref{fig:event_1_lag_maps}, which was obtained with the peak cross-correlation method. To use phase maps to recover time lag between channels allows to discard contribution of background and foreground. Moreover, we extract the signal of the pulsating loops pixel by pixel without any assumption on the shape of the structure.

	\begin{table*}
                \centering
                        \caption{Comparison of time lag values between the two methods: peak cross-correlation values and differences of phase. These are the averaged time lag in minutes in the black contour.}
                       \label{table:event_1_time_lag_contour}
                        \begin{tabular}{c c c}
         		       \hline\hline
                                Pairs of channels &  Time lag (min) &  Time lag (min) \\
                                \, & from cross-correlations (Fig.~\ref{fig:event_1_lag_maps}) & from differences of phase Fig~.(\ref{fig:event_1_phase_maps_lag})  \\
                                \hline

                                335-211 & 113 & 122 \\
                                211-193 & 26 & 20 \\   
                                335-193 & 145 & 142 \\                                
                                94-335 & -115 & -114 \\
                                335-171 & 142 & 155 \\
                                171-131 & -1 & -51 \\                                

                                \hline
                        \end{tabular}
	\end{table*}

\paragraph{}We compare time lags averaged over the contour obtained with the peak cross-correlation method and with the phase difference method in Table.~\ref{table:event_1_time_lag_contour}. 
With both methods we find similar time lags and approximately the same ordering of channels. 
For method 1 (cross-correlations), the time lag for 335-193 is larger than the time lag for 335-171 (145 min compared to 142 min). So the intensity peaks first in the 171 channel and then in the 193 channel. For the method 2 (phase difference), the time lag for 335-193 is inferior to the time lag for 335-171 (142 min compared to 155 min). The second method gives an ordering of channel that is more conform to the one predicted from a monotonic temperature decrease and from the AIA response functions. Only the 94-335 and 171-131 pairs give a time lag that is not expected in this predicted ordering. Indeed, the absolute value of the time lag of 94-335 (114 min) is smaller than the one of 335-211 (122 min) and thus the intensity peaks in 94 before that in 211. Similar ordering is found with method 1 as the time lags for 94-335 (115 min) and 335-211 (113 min) are the same given the time resolution. For the 171-131 pair the time lag is -51 min with method 2 compared to -1 min with method 1.
Method 2 (phase difference) is limited in some passbands by the Fourier power spectrum at the detected frequency: as the normalized power is weaker for the 171 and the 131 channels, method 2 is more affected by noise and thus less reliable for the 171-131 pair.

    \section{Discussion}\label{sec:discussion}   
     
\paragraph{}We selected three cases of long-period intensity pulsations and analyzed them in the six coronal channels of AIA. These pulsations have periods of several hours, from 3.8 hours to 9.0 hours for the selected events. They are strongly visible in most of the passbands and are clearly matching loops structures in active regions. In each case, only some loops among those of the active regions are pulsating. 

        \subsection{Evolution of the thermal structure of pulsating loops} 
        
\paragraph{}Multi-spectral analysis with AIA allows to proceed with a thermal study of these cases. The analysis of the thermal structure, i.e. the distribution and evolution of temperatures and densities, of the loops where pulsations occur, reveals periodic variations. The DEM analysis also reveals high amplitude variations in the coolward wing of the DEM. We link these variations to a signature of heating of these loops.
The variation of the DEM slope, which are large, can tell us about the timescales of the heating. Even though \citet{guennou2013} have shown that the DEM slope is poorly constrained, we interpret here only the relative variations of the slope. A steep slope implies that the heated plasma does not have enough time to cool down to lower temperatures, and then that the frequency of individual heating events is high compared to the cooling time.
On the contrary, when the slope is weak, a greater quantity of plasma has cooled down before being re-heated, implying a lower frequency of heating.

There are clear pulsations for the total emission measure for the three events studied. For event~1, there is a peak of normalized power above the adopted threshold (99\% of confidence level) for the peak temperature. For event~2 and event~3, there are peaks of normalized power, respectively for the peak temperature and the slope, below this threshold but still at high confidence levels. These small peaks have respectively 97\% (8.7\(\sigma\)) and 96\% (8.4\(\sigma\)) of confidence level. Pulsations do not come out clearly from the Fourier analysis of the slope and the peak temperature.
This can be explained by several effects impacting the DEM results:

First, we remind that the DEM analysis is made without any subtraction of the background and foreground emissions because it would not be practical nor meaningful given that the detected pulsations are much longer than the typical life time of loops. However background emissions can be seen in images presented in Fig.~\ref{fig:event_1_images}, Fig.~\ref{fig:event_2_images}, and Fig.~\ref{fig:event_3_images}, especially at 131~\AA, where we can clearly see the emission from the transition region. Superimposition of structures with different temperature and density conditions can partially hide the signal from the pulsating loops.

Second, the importance of the background and foreground emission can be different in the different passbands. As an illustration, for event~1, where there are no pulsations (given the \(99\%\) confidence level) for the DEM slope, the normalized pulsating power is the strongest in the hotter channels (193, 211 and 335). The three other ones which could constrain the cool wing of the DEM may be the most affected by the background and foreground emission. On the contrary, for event~3, pulsations are exclusively visible in the cooler passbands: 94 (i.e. the cool component of 94), 131, and 171.

Third, \citet{guennou2013} showed that the slope is poorly constrained and thus sensitive to random variations in the light curves, which may explain why the time series of the slope is itself noisy.

To conclude, the lack of clear pulsations (given the \(99\%\) confidence level) for the DEM slope for the three events and for the DEM peak temperature for event~2 and event~3, could come from a combination of the three above described effects. In addition even if it does not come out clearly from the Fourier spectrum, there are indications of periodic pulsations in the DEM slope from correlations analysis. Therefore we conclude that there are oscillations in the thermal structure of these loops.

        \subsection{Widespread cooling} 
\paragraph{}The studied loops are seen in a state of cooling, as was shown using the same method as V\&K12: time lag maps for six pairs of channels show widespread cooling for the three active regions, independently of the pulsating behavior of some of the loops in these active regions. To avoid the effect of the background and foreground emissions, we confirm this result using a second method, by extracting the pulsating component of the signal in Fourier space.

But why do we only observe cooling? The usual explanation is that the density is too small for the plasma to be detectable during the heating phase. The widespread cooling can also be a signature of heating mechanisms: V\&K12 and \citet{viall&klimchuk2013} interpret it as a signature of heating by nanoflare storms.
In the nanoflare model indeed, loops strands are heated (by impulsive events) and cool independently \citep[e.g][]{warren_2003}. In that case, the cooling time of a bundle can be much longer than the typical life time of individual loops. Nevertheless, for \citet{lionello_2013}, this widespread cooling is also consistent with quasi-continuous heating, as their model can reproduce long cooling times: they can be also interpreted by superimpositions of loops cooling independently along the line of sight and are not necessarily a signature of impulsive heating mechanisms.

In the time lag maps we do not notice differences between the pulsating loops and the rest of the active regions. The cooling phase is thus the same for the entire active region but that does not exclude different processes of heating between the pulsating loops and the rest of the active regions. The heating mechanism may or may not be the same inside and outside of the pulsating region but there is not yet enough information to conclude.

        \subsection{Possible physical explanations for long-period pulsations} \label{sec:discussion_3}        
\paragraph{}In order to explain long-period pulsations seen in the coronal EUV channels and in the thermal structure of warm loops, we can put forward at least two physical explanations depending of the supposed nature of the heating.
If it is impulsive, we could envisage periodic nanoflare storms, i.e. nanoflare storms with periodical changes of the frequency of individual heating events, from high frequency (short time between nanoflare events) to a lower frequency (more time for the plasma to cool down before being re-heated). Then the DEM would change according to the changes in heating frequency. A change in the frequency of nanoflares might come from a coupling between the motions at the footpoints  and the magnetic structure of the loops. 

\begin{figure}[h!]
		\centering
                 \resizebox{\hsize}{!}{\includegraphics{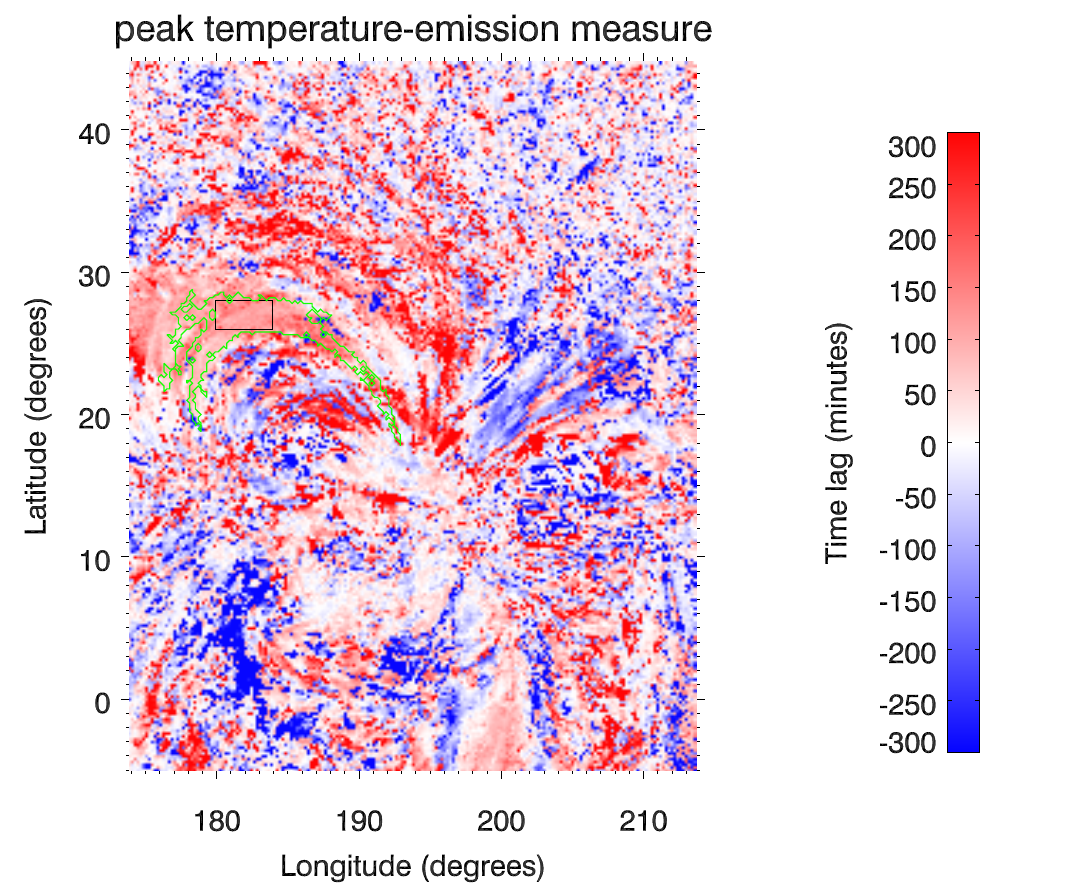}}
                 \caption{Time lag map obtained from differences of phase maps of the peak temperature and of the emission measure. We explored time shifts from -300 minutes to 300 minutes. The green contour is the detected contour and the black contour is the contour manually selected.}
                 \label{fig:dem_time_lag_event_1}
	\end{figure}

On the other-hand, a quasi-continuous heating at the loops footpoints would also be a good candidate to explain our observations. This scenario drives a condensation-evaporation cycle (thermal non-equilibrium): the plasma condenses at the top of the loops, this material falls toward the footpoints, then chromospheric material from the footpoints evaporates and fills the loops with hot plasma. Actually, even if the role of thermal non-equilibrium was often ruled out in the heating of warm coronal loops \citep{klimchuk_2010}, we find that the long-period intensity pulsations observed in loops are consistent with condensation-evaporation cycles.
Fig.~\ref{fig:dem_time_lag_event_1} represents the time lag between the peak temperature and the emission measure for the entire field of view for event~1. This map is built by subtracting phase maps of the DEM peak temperature and of the emission measure at the main frequency detected for event~1. This allows us to obtain the time lag map between the peak temperature and the emission measure for only the periodic component of these parameters. Therefore, pulsating loops at the detected frequency are isolated in Fourier space. This map shows a time lag of 123~min on average in the selected contour which is what we already found in Fig.~\ref{fig:correlations_dem_event_1} without Fourier filtering. This corresponds to a de-phasing of 81\(\degree\). This result is to be compared with Fig.~6 in \citet{mikic_2013}, where they compare the evolution of the temperature and the density at the apex of a symmetric simulated loop, with a nonuniform cross-section area and a nonuniform heating. The temperature and the density of their simulation have cyclic variations with a period of 5.0 hours, which is at the order of the periods that we observed. The density curves have a delay of about 1.5 hours compared to the temperature curve. That implies that the temperature variations always precede the density variations with a de-phasing of approximately 108\(\degree\). For this simulation the authors note development of incomplete condensation cycles localized near the loop legs and a temperature at the loop apex that remains above 1~MK. The observed and modeled de-phasings are comparable, as far as \citet{mikic_2013} use the density and we have only access to the total emission measure.
        
 \paragraph{}Even if long-period intensity pulsations are connected to heating processes in loops, these pulsations occur only in some loops of the studied active regions and not in the entire active regions. In other words, why don't we have long-period intensity pulsations in all loops? This could be due to the background and foreground emission as developed earlier in the text. Intrinsic properties of loops could also explain why some loops do not show this kind of pulsations. For \citet{mikic_2013}, the plasma response to a steady heating mainly concentrated at the loops footpoints depends on the loops geometry: the uniformity and the symmetry of the cross-sectional area, the symmetry of the loop profile and the symmetry of the heating profile with different heating functions. The geometry of the loops and of the heating could lead to different condensation strengths that they call "incomplete" and "complete" condensation. This connects thermal non-equilibrium processes to both coronal rain and heating of warm loops.
 
The wide distribution of observed periods could be due to inherent properties of the loops. In \citet{auchere2014} a possible relation between the length of the loops and the period of the pulsations was discussed.  Here also we notice that for these three events, the longer the period of pulsations, the larger the detection area. It remains to be confirmed that the loop length is a good candidate to explain the different periods observed in pulsating loops.

    \section{Summary and Conclusions}  
\paragraph{}\citet{auchere2014} have shown that long-period intensity pulsations are a common phenomenon in coronal loops. Using the same detection algorithm, we have detected many other cases in the six coronal channels of AIA. Among these, we selected three events detected in three different active regions. The observed pulsations have respectively periods of 9.0 hours, 5.6 hours, and 3.8 hours. The normalized power maps of the three studied active regions clearly show that these pulsations are strongly localized in some loops, but not in all loops, of active regions. Thus, either the occurrence of these cycles depends on specific loop properties or some events could be masked by the background and foreground emissions.

We investigate the physical properties underlying these pulsations. First by means of a DEM analysis using the active region DEM model from \citet{guennou2013}. We notice periodic variations of the shape of the DEM during the sequences studied. The thermal structure of these loops changes periodically from a strong multithermal structure to a more isothermal structure. We link these variations to a signature of heating mechanisms. Moreover, using the same method than in V\&K12, we notice that the studied active regions are seen mainly in their cooling phase. This is confirmed by extracting in Fourier space the loops at the detected frequency, allowing to remove contributions from the background and foreground emissions. Although this widespread cooling could be linked to nanoflare heating, it does not inevitably imply impulsive heating. In fact, \citet{lionello_2013} shows, by means of 3D-hydrodynamic simulations, that thermal non-equilibrium processes could be also consistent with widespread cooling. Our observations bring a new element to this debate. Since these pulsations appear to be common in coronal loops and are linked, as we suggest, to the heating of the plasma, then any model of loop heating must be able to reproduce these new observations.

\acknowledgements
The SDO/AIA images are available by courtesy of NASA/SDO and the AIA science teams. This work made extensive use of the AIA archive at MEDOC, http://medoc-sdo.ias.u-psud.fr

\appendix

\section{Event 2}
\label{sec:event2_figures}

     \subsection{Data sample}      
\begin{figure*}
		\centering
                 \includegraphics[width=\linewidth, trim = 0 0 0 2.05cm, clip]{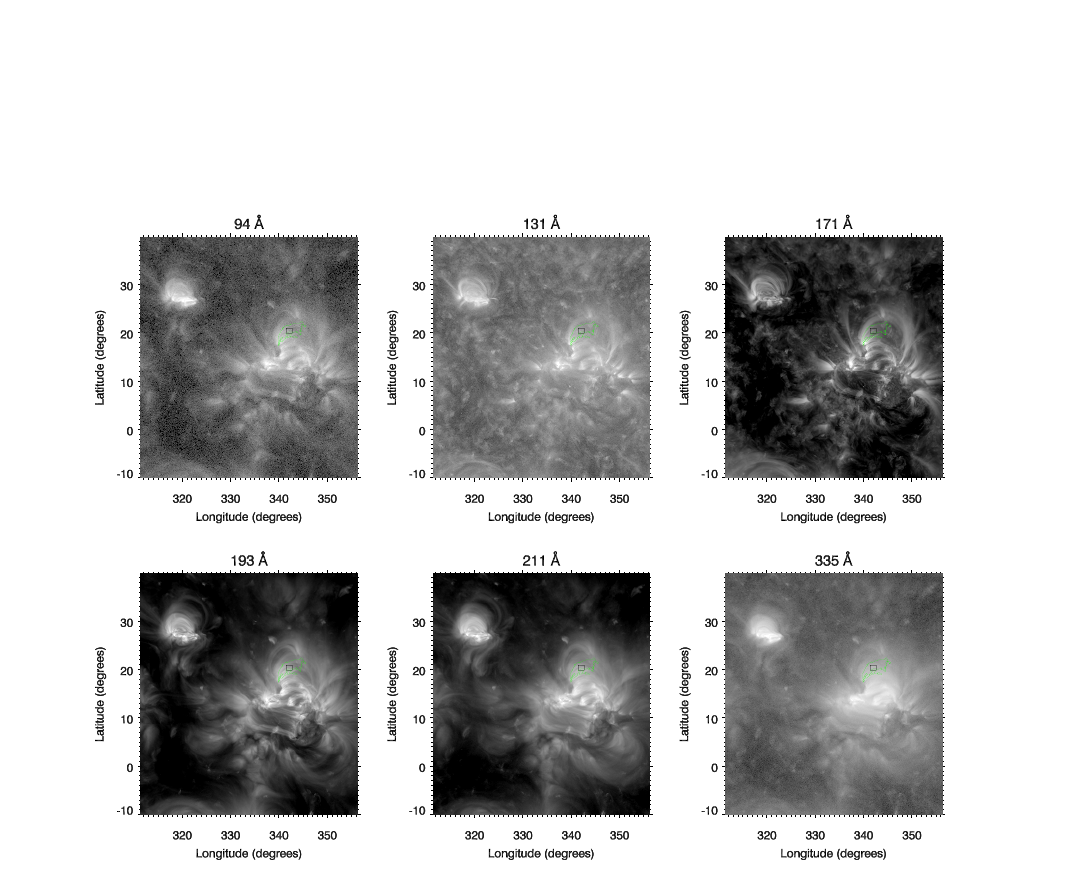}
                 \caption{Same as Fig.~\ref{fig:event_1_images} for event~2 on December 30, 2012 at 21:28~UT. This event is localized in NOAA AR 11637.}
                 \label{fig:event_2_images}            
	\end{figure*}

\begin{figure*}
		\centering
                 \includegraphics[width=\linewidth]{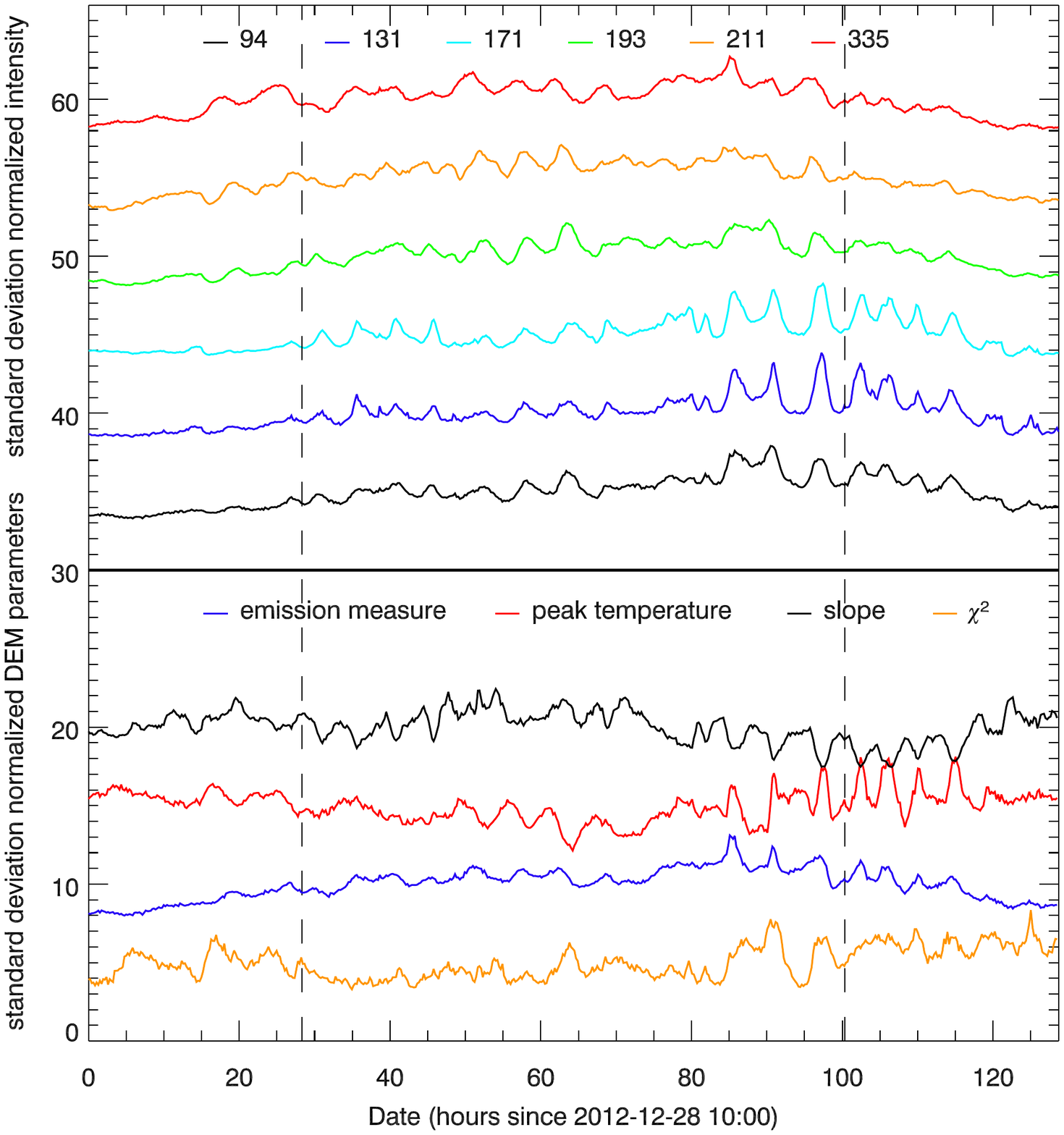}
                 \caption{Same as Fig.~\ref{fig:light_curves_1} for event~2 from December 28, 2012 10:00~UT to January 02, 2013 18:42~UT. }
                 \label{fig:light_curves_2}
	\end{figure*}

\paragraph{}Fig.~\ref{fig:event_2_images} shows images of a second exemple of pulsating loops with long-period on December 30, 2012 at 21:28~UT. For this event~2, the NOAA AR 11637 was tracked from December 28, 2012 10:00~UT to January 02, 2013 18:42~UT (i.e. about 129 hours). As for event~1, we focused on the three middle days of this sequence, from December 29, 2012 14:26~UT to January 01, 2013 14:29~UT (i.e. 72 hours). The contour in green is the contour detected in the 171~\AA~channel centered at 342.5\(\degree\) of longitude and 20.5\(\degree\) of latitude with an area of 12.9 heliographic square degrees (1495 Mm\(^{2}\)). We manually selected a small contour (in black) included in the detected contour, to obtain the time series presented in Fig.~\ref{fig:light_curves_2}.

	\begin{figure*}
		\centering
                 \includegraphics[width=\linewidth]{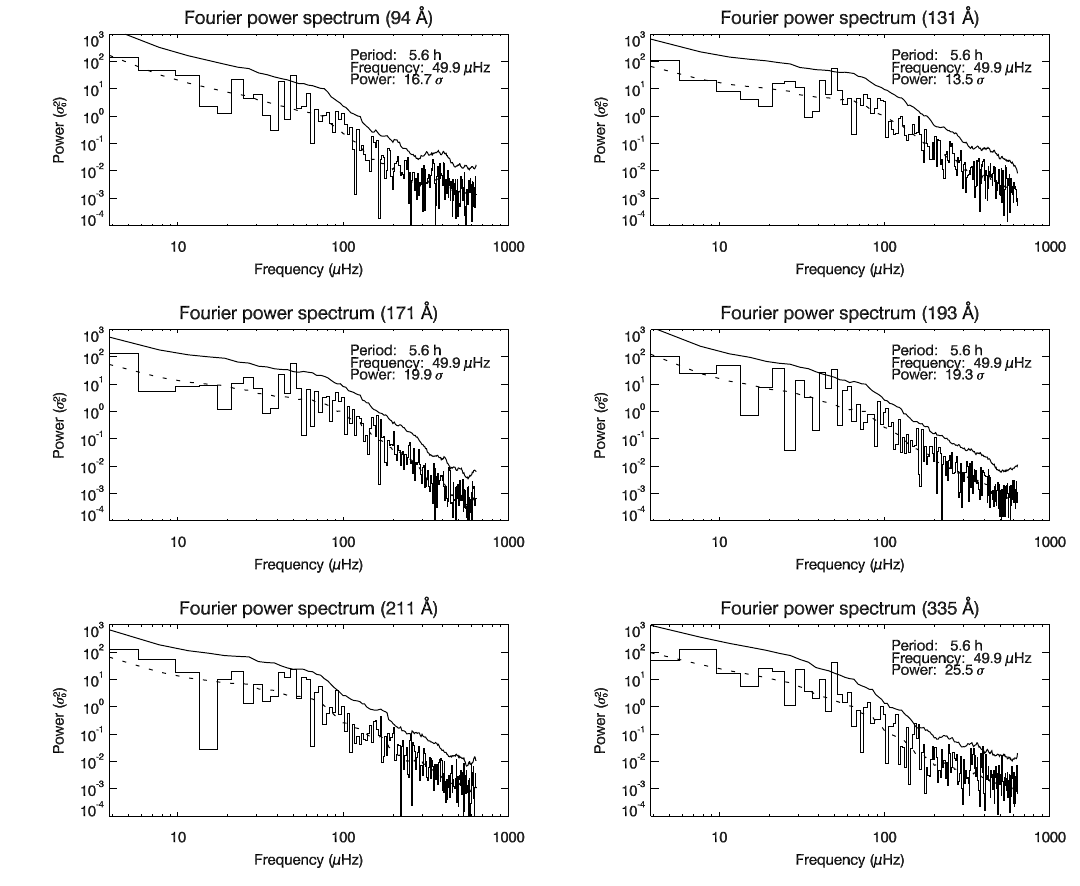}
                 \caption{Fourier power spectra for event~2, same as Fig.~\ref{fig:event_1_spectrum}}
                 \label{fig:event_2_spectrum}
	\end{figure*}

We display the normalized Fourier spectra in Fig.~\ref{fig:event_2_spectrum}. In the 94~\AA, 131~\AA, 171~\AA, 193~\AA, and 335~\AA~channels there is a significant peak of power at 49.9~\(\mu\)Hz (5.6 hours). The normalized powers are respectively 16.7\(\sigma\), 13.5\(\sigma\), 19.9\(\sigma\), 19.3\(\sigma\), and 25.5\(\sigma\). Even if there is no peak above 10\(\sigma\) in 211, we can notice a small peak with a normalized power of 9.6\(\sigma\) at 49.9~\(\mu\)Hz. At 94~\AA, and 193~\AA~there is also a small peak at 42.2~\(\mu\)Hz with respectively 8.1\(\sigma\) and 11.7\(\sigma\) of normalized power.

\begin{figure*}
		\centering
                 \includegraphics[width=\linewidth, trim = 0 0 0 2.05cm, clip]{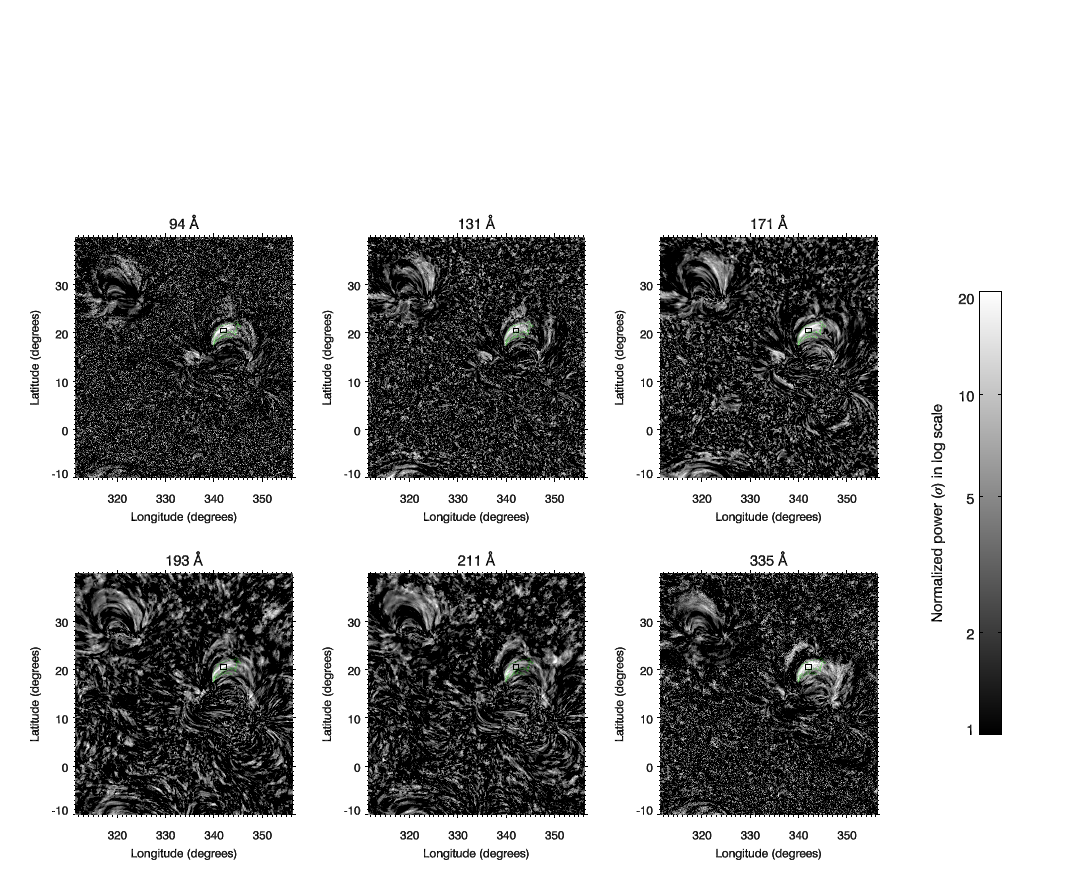}
                 \caption{Same as Fig.~\ref{fig:event_1_power_maps} with the same scale. Maps of normalized power at the frequency of detection (49.9~\(\mu\)Hz, i.e. 5.6 hours) for event~2. The black contour, included in the detected contour at 171~\AA~(in green), is the contour manually selected for detailed time series.}
                 \label{fig:event_2_power_maps}
	\end{figure*}

Fig.~\ref{fig:event_2_power_maps} represents the maps of normalized power for event~2, at the detection frequency (49.9~\(\mu\)Hz, i.e. 5.6 hours). Loops of NOAA AR 11637 are clearly seen in these maps for all the passbands. The averaged normalized power is larger than 10\(\sigma\) for all the passbands, except 211~\AA~(7\(\sigma\)). Loops of NOAA AR 11640, a second active region around 320\(\degree\) in longitude and 30\(\degree\) in latitude, have also normalized power above 10\(\sigma\). The automatic algorithm does not detect these loops with the set of arbitrary thresholds. This suggests that more of these events are probably not detected by our code. Loops of this second active region (emerging at the beginning of the sequence) seems to have almost the same length, that can be a reason why they have the same frequency of pulsations (see section~\ref{sec:discussion_3}). 

      \subsection{Differential emission measure (DEM) analysis}   
\begin{figure*}
		\centering
                 \includegraphics[width=\linewidth]{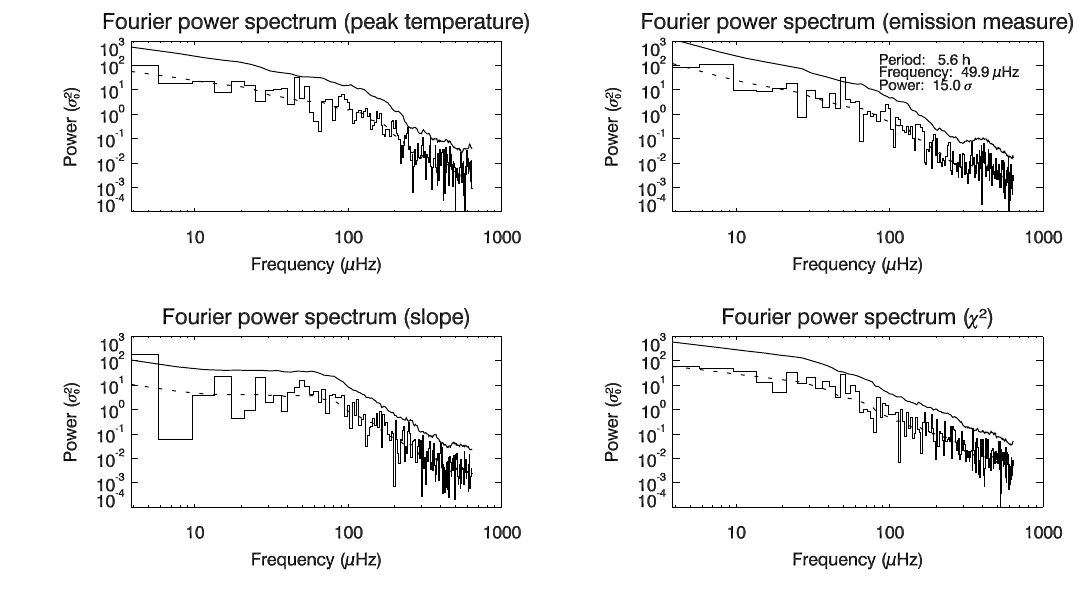}
                 \caption{Fourier power spectra for the three parameters of the DEM model for event 2, same as Fig.~\ref{fig:dem_event_1_spectrum}}
                 \label{fig:dem_event_2_spectrum}
	\end{figure*}

\paragraph{} The evolution of the mean DEM slope, peak temperature, emission measure, and \(\chi^2\) in the selected contour of event~2 is presented in Fig.~\ref{fig:light_curves_2}.
Fig.~\ref{fig:dem_event_2_spectrum} represents the power spectra of these parameters. For the emission measure there is a peak power at 49.9~\(\mu\)Hz, i.e. 5.6 hours. The power at the peak is 15.0\(\sigma\). If we look under the detection threshold at 49.9~\(\mu\)Hz, there is a peak with a power of 8.0\(\sigma\) for \(\chi^{2}\), but its variations from to 0.3 to 1.8 indicates a good fit.
For the peak temperature, there is a small peak (8.7\(\sigma\)) at 46.1~\(\mu\)Hz. For the slope there is no significant peak of power. 

\begin{figure}
		\centering
		\includegraphics[width=0.5\textwidth]{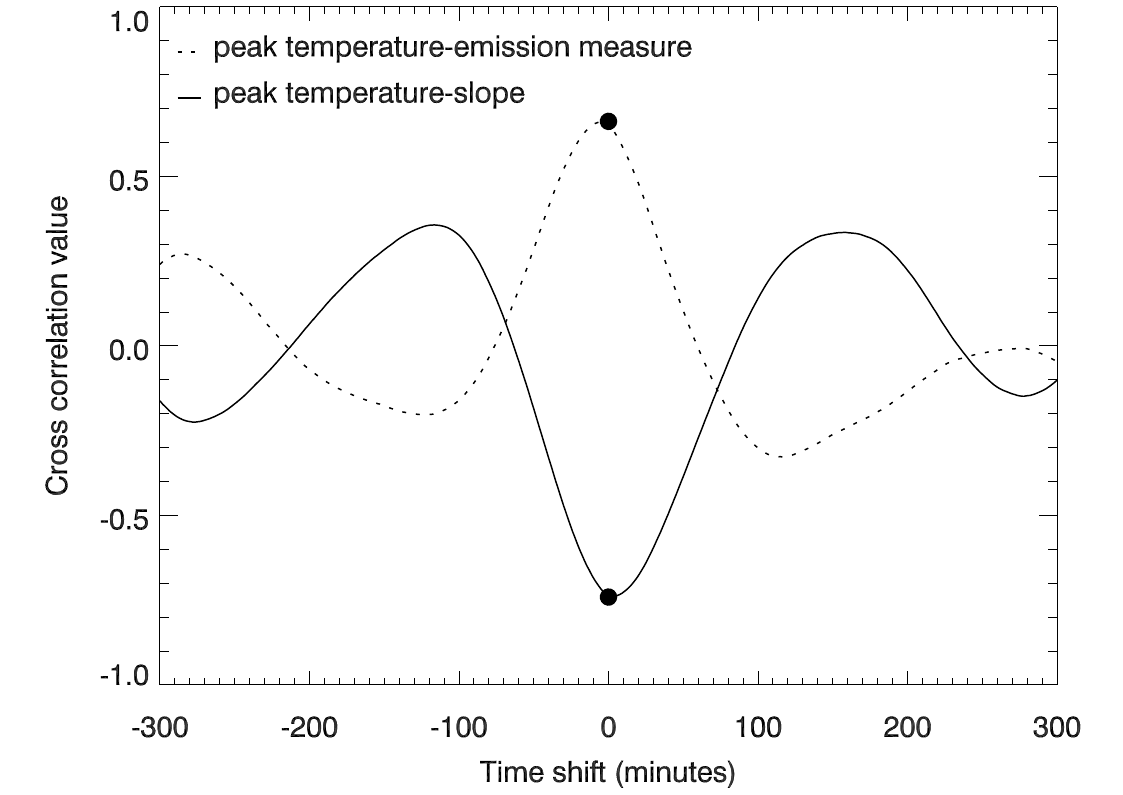}
		\caption{Same as Fig.~\ref{fig:correlations_dem_event_1}. We used the averaged curves at 1 min of cadence averaged over the small black contour presented in Fig.~\ref{fig:event_1_images} between January 03, 2013 22:28~UT (97 h after the beginning of the long sequence) and January 04, 2013 21:28~UT (120 h after the beginning of the long sequence).}
		\label{fig:correlations_dem_event_2}
	\end{figure}

Even if the power is weak for the peak temperature and there are no pulsations found with the Fourier analysis for the slope, we found a clear correlation between the slope and both the peak temperature and the emission measure, between 97 hours and 120 hours after the beginning of the full sequence. In Fig.~\ref{fig:correlations_dem_event_2}, we plot the cross-correlation values for DEM peak temperature vs.\ emission measure and slope. We find a clear correlation between the peak temperature and the emission measure (cross-correlation value of 0.66) and anti-correlation between the peak temperature and the slope (cross-correlation value of -0.74). We can conclude that there are indications that the slope varies periodically with periods of hours and thus that the thermal structure of the studied loops undergoes periodical changes.
      
The amplitudes of variations of the DEM parameters are larger than for event~1: there are variations from 2.2 to 3.2~MK (a relative amplitude of 45\%) for peak temperature, from \(3.2 \times 10^{27}\) to \(6.4 \times 10^{27}\)~cm\(^{-5}\) for emission measure (a relative amplitude of 97\%), and from 1.7 to 4.8 for the slope (a relative amplitude of 176\%).

     \subsection{Evidence for widespread cooling}   
  
	\begin{figure*}
		\centering
                 \includegraphics[width=\linewidth, trim = 0 0 0 2.05cm, clip]{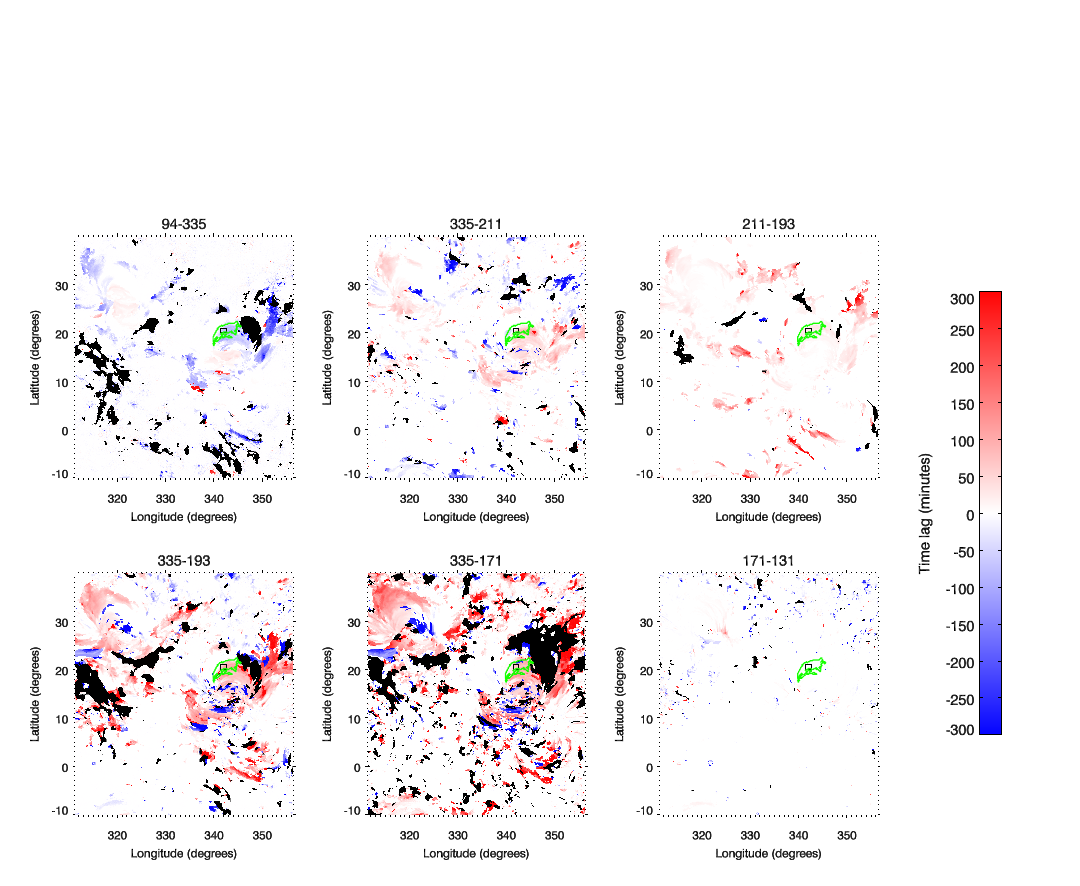}
                 \caption{Same as Fig.~\ref{fig:event_1_lag_maps} for event~2 with the same scale.}
                 \label{fig:event_2_lag_maps}
	\end{figure*}
        
\begin{figure*}
		\centering
		\includegraphics[width=\linewidth, trim = 0 0 0 2.05cm, clip]{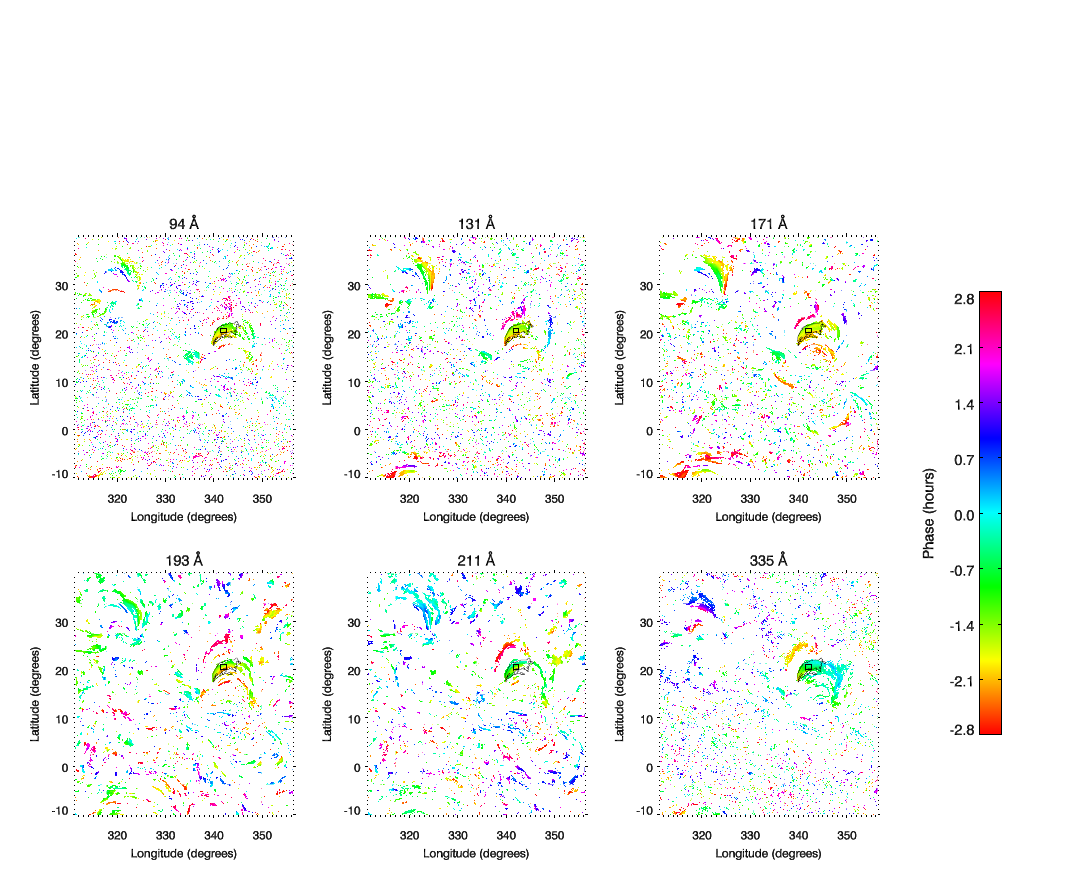}
                 \caption{Same as Fig.~\ref{fig:event_1_phase_maps} for event~2 with the same scale.}
                 \label{fig:event_2_phase_maps}
	\end{figure*}

\begin{figure*}
		\centering
                 \includegraphics[width=\linewidth, trim = 0 0 0 2.05cm, clip]{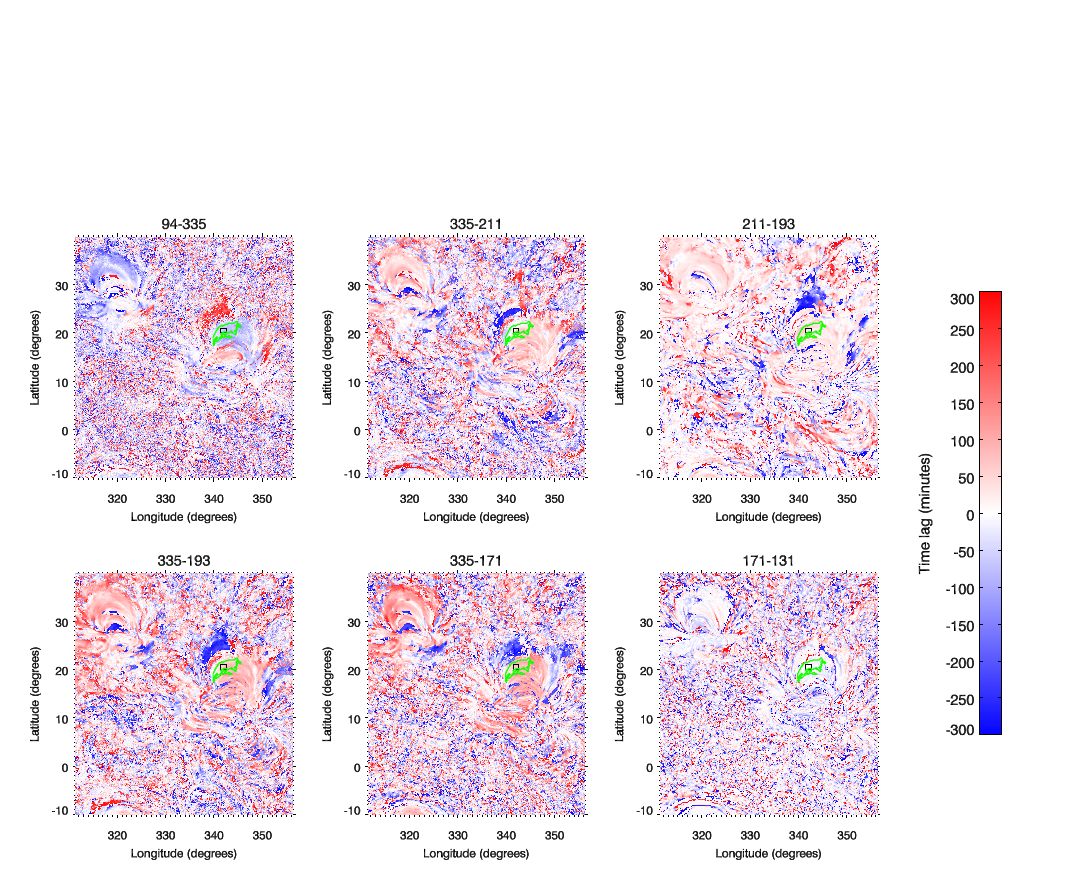}
                 \caption{Same as Fig.~\ref{fig:event_1_phase_maps_lag} for event~2 with the same scale.}
                 \label{fig:event_2_phase_maps_lag}
	\end{figure*}

\paragraph{}Fig.~\ref{fig:event_2_lag_maps} represents the time lag maps made with the peak cross correlation values. Fig.~\ref{fig:event_2_phase_maps} represents the phase maps at 49.9~\(\mu\)Hz. For the two active regions included in the field of view, the phases are constant along the loops (as seen in Fig.~\ref{fig:event_1_phase_maps} for event~1) but there is a gradient of phases across the bundle of loops. Further analysis is needed to explain this gradient. Fig.~\ref{fig:event_2_phase_maps_lag} shows the differences of phase maps for six pairs of channels. 
We included the detected contour in green and the contour manually selected in black. On most of the field of view the plasma is in a state of cooling, for both Fig.~\ref{fig:event_2_lag_maps} and Fig.~\ref{fig:event_2_phase_maps_lag}: time lags are mainly positive for the 335-211, 211-193, 335-211, 335-193, and 335-171 pairs , mainly negative for 94-335 and equal to zero for 171-131. The mean time lags in the black contour can be compared in Table~\ref{table:event_2_time_lag_contour}. Using the cross-correlation method, intensity peaks in the following channel order: 335, 211, 171 and 131, 94, and 193. Using the phase difference method, the order becomes: 335, 211, 193, 94, 171, and 131, that is the order expected from the expected temperature evolution and from the AIA response functions. This second method is better in this case (event~2) because the power is high enough in all the passbands at the detected frequency. However, for pairs like 94-335 and 335-193 time lags are almost equal.
We can notice that these time lags are smaller than the time lags found for event~1 (about the half). These smaller time lags are likely to be due to the smaller loop length (smaller detection area) for event~2. 

	\begin{table*}
		\centering
                 \caption{Same as Table~\ref{table:event_1_time_lag_contour} for event~2.}
                 \label{table:event_2_time_lag_contour}
                        \begin{tabular}{c c c}
                                \hline\hline
                                Pairs of channels &  Time lag (min) &  Time lag (min) \\
                                \, & from cross-correlations (Fig.~\ref{fig:event_2_lag_maps}) & from differences of phase Fig~.(\ref{fig:event_2_phase_maps_lag})  \\
                                \hline

                                335-211 & 35 & 37  \\
                                211-193 & 25 & 31 \\
                                335-193 & 80 & 68 \\
                                94-335 & -71 & -68 \\                                
                                335-171 & 69 & 73 \\
                                171-131 & 1 & -4 \\
                                
                                \hline
                        \end{tabular}
	\end{table*}    

\section{Event 3}
\label{sec:event3_figures}
     \subsection{Data sample}   
\begin{figure*}
		\centering
                 \includegraphics[width=\linewidth, trim = 0 0 0 2.05cm, clip]{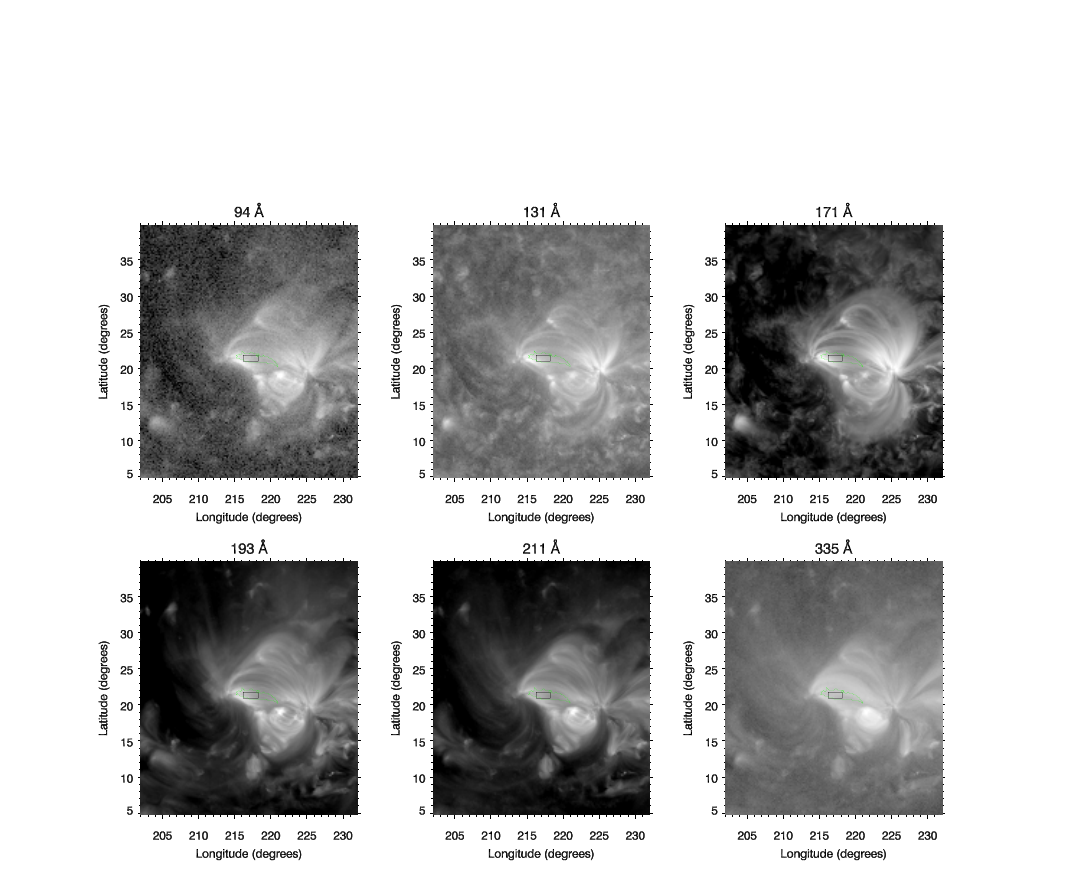}
                 \caption{Same as Fig.~\ref{fig:event_1_images} for event~3 on August 09, 2011 12:12~UT. This event is localized in NOAA  AR 11268.}
                 \label{fig:event_3_images}
	\end{figure*}

\begin{figure*}
		\centering
                 \includegraphics[width=\linewidth]{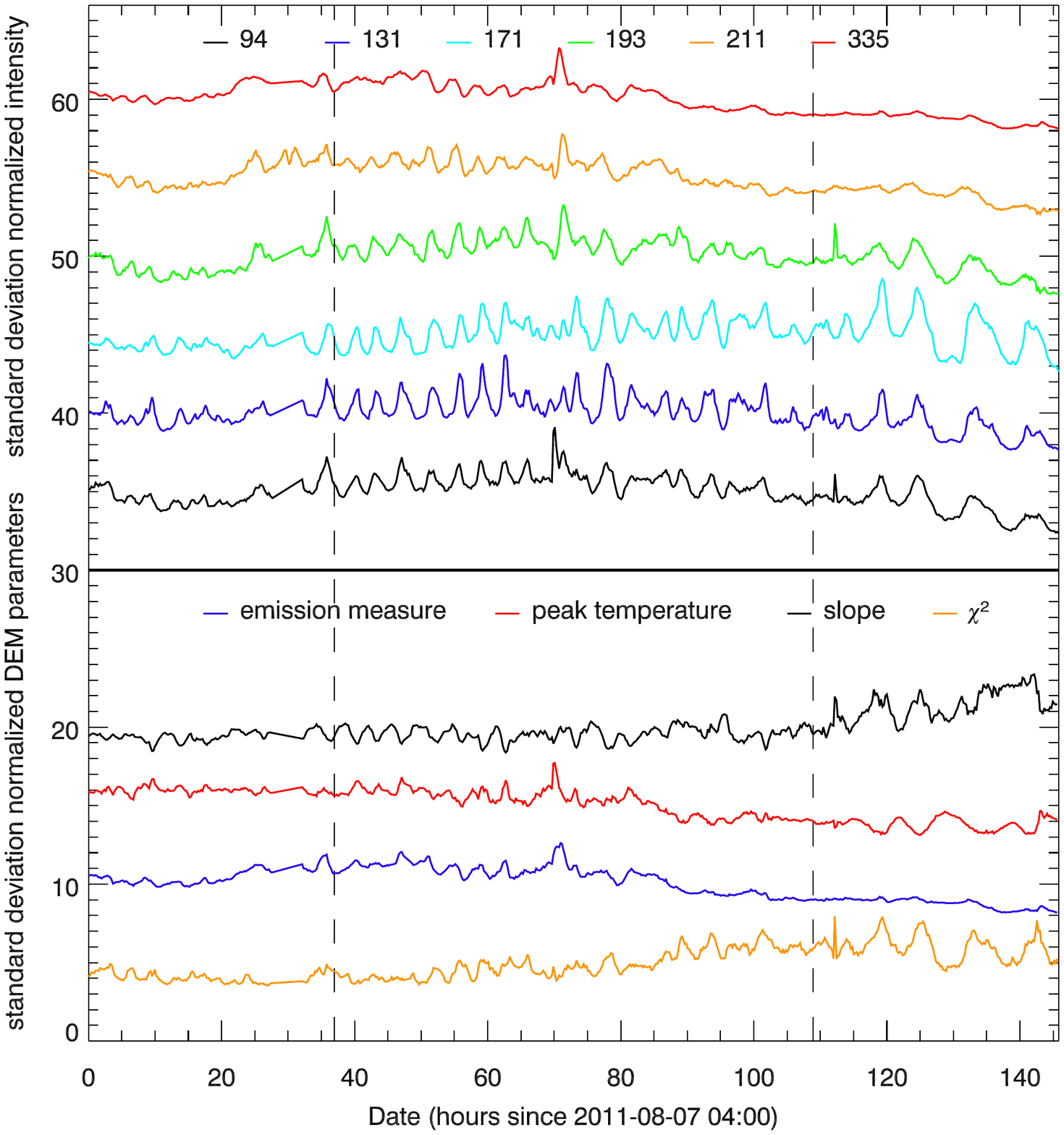}
                 \caption{Same as Fig.~\ref{fig:light_curves_1} for event~3 from August 07, 2011 04:00 UT to August 13, 2011 05:49 UT. }
                 \label{fig:light_curves_3}
	\end{figure*}

\begin{figure*}
		\centering
                 \includegraphics[width=\linewidth]{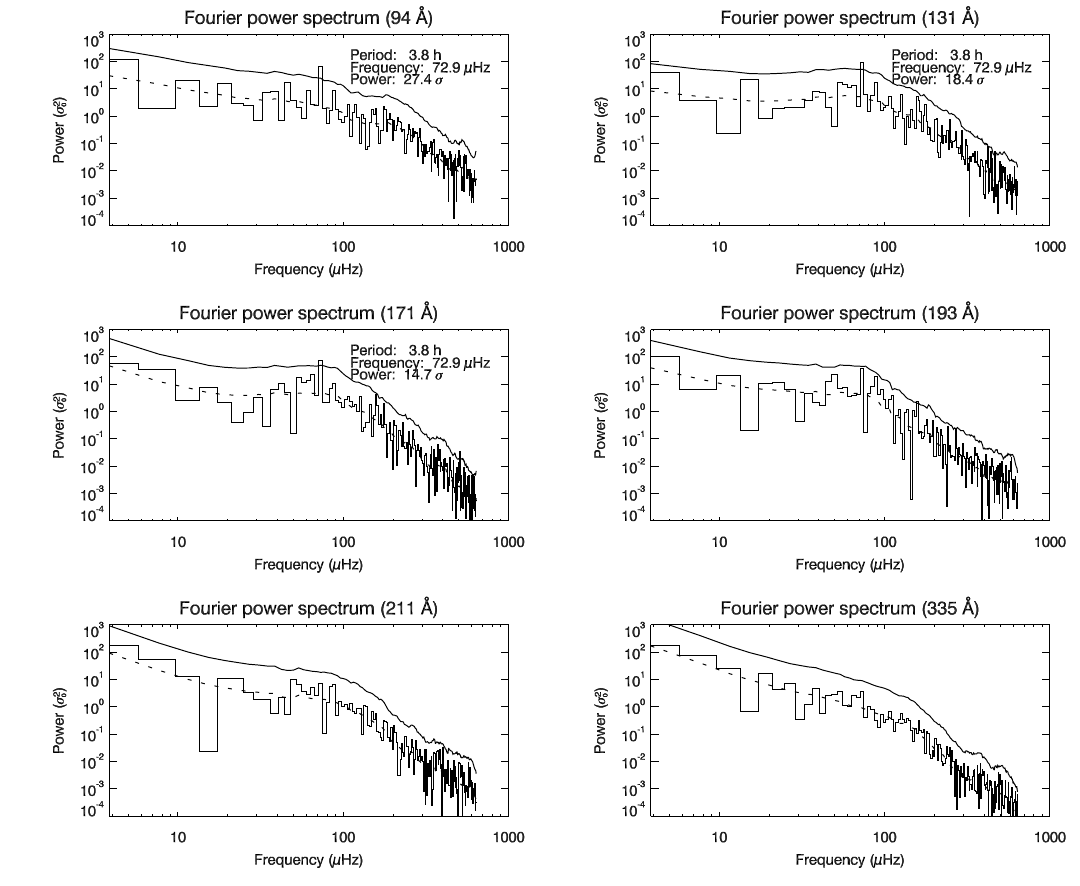}
                 \caption{Fourier power spectra for event 3, same as Fig.~\ref{fig:event_1_spectrum}}
                 \label{fig:event_3_spectrum}
	\end{figure*}

\paragraph{}Fig.~\ref{fig:event_3_images} shows images of a last example, event~3, on August 09, 2011 12:12~UT. For this event, the ROI (45\(\degree\) in longitude)  was tracked from August 07, 2011 04:00~UT to August 13, 2011 05:49~UT (i.e. about 146 hours), and we select the three middle days of this sequence, from August 08, 2011 16:53~UT to August 11, 2011 16:56~UT (i.e. 72 hours). Pulsations are detected in loops located above the core of NOAA AR 11268. The detected contour at 94~\AA~(in green) is centered on 218.0\(\degree\) of longitude and 21.4\(\degree\) of latitude with an area of 6.3 heliographic square degrees (730 Mm\(^{2}\)). In Fig.\ref{fig:light_curves_3}, we represent the average light curves in the selected contour (in black). The corresponding Fourier spectra in Fig.~\ref{fig:event_3_spectrum} show strong signal in the 94~\AA, 131~\AA, and 171~\AA~channels. Pulsations are detected in these three bands at 72.9~\(\mu\)Hz (3.8 hours) with respectively a power of 27.4\(\sigma\), 18.4\(\sigma\), and 14.7\(\sigma\) above the estimated background spectrum. At 211~\AA~and 335~\AA~there is no significant power peak, but at 193~\AA~there is a small peak at 8.2\(\sigma\). In this last channel, there is also a small peak (7.0\(\sigma\)) at 88.3~\(\mu\)Hz. 
                
\begin{figure*}
		\centering
                 \includegraphics[width=\linewidth, trim = 0 0 0 2.05cm, clip]{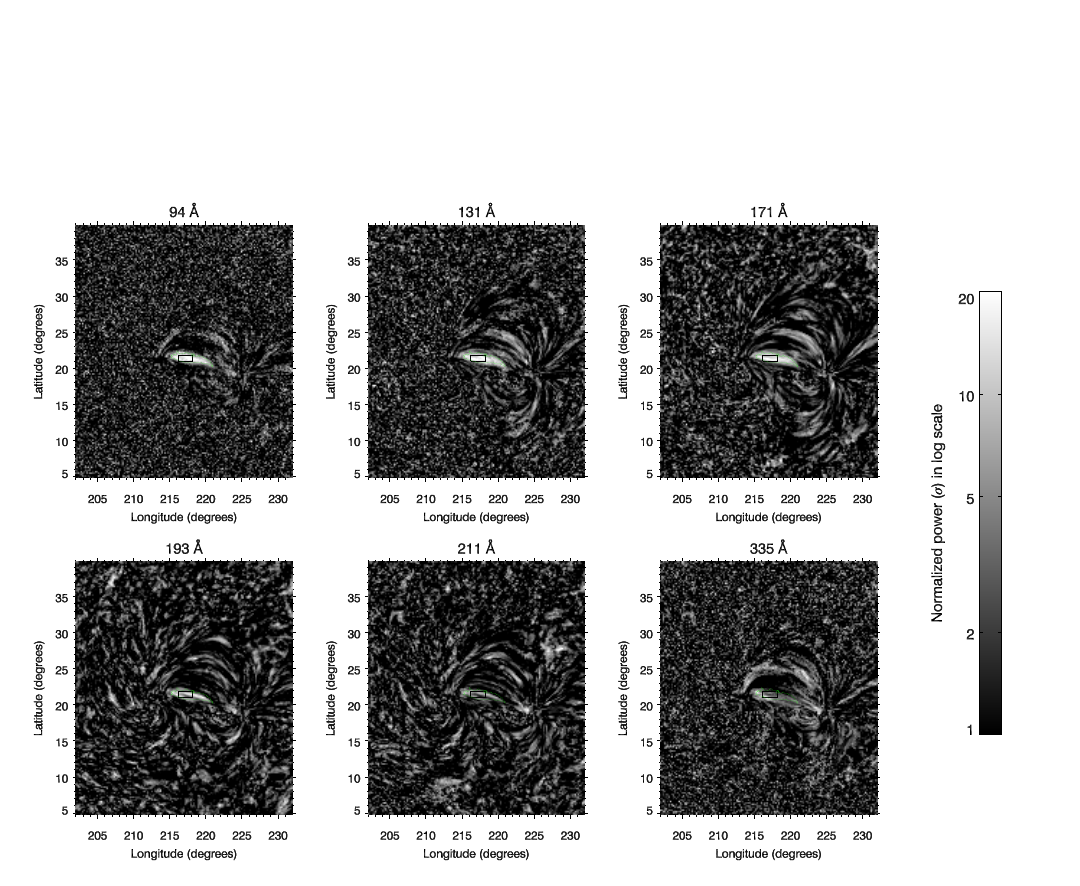}
                 \caption{Same as Fig.~\ref{fig:event_1_power_maps} with the same scale. Maps of normalized power at the frequency of detection (72.9~\(\mu\)Hz, i.e. 3.8 hours) for event~3. The black contour, included in the detected contour at 94~\AA~(in green), is the contour manually selected for detailed time series.}
                 \label{fig:event_3_power_maps}
	\end{figure*}

Fig.~\ref{fig:event_3_power_maps} represents the maps of normalized power for event~3, at the detection frequency (72.9~\(\mu\)Hz, i.e. 3.8 hours). The normalized power in the 94~\AA, 131~\AA, and 171~\AA~channels is higher than 14\(\sigma\) on average inside the black contour. For the other passbands, the normalized power is weaker, but the power maps also display loops near the core of the active region.

     \subsection{Differential emission measure (DEM) analysis}      

\paragraph{} The evolution of the mean DEM slope, peak temperature, emission measure, and \(\chi^2\) in the selected contour for event~3 is presented in Fig.~\ref{fig:light_curves_3}.

\begin{figure*}
		\centering
                 \includegraphics[width=\linewidth]{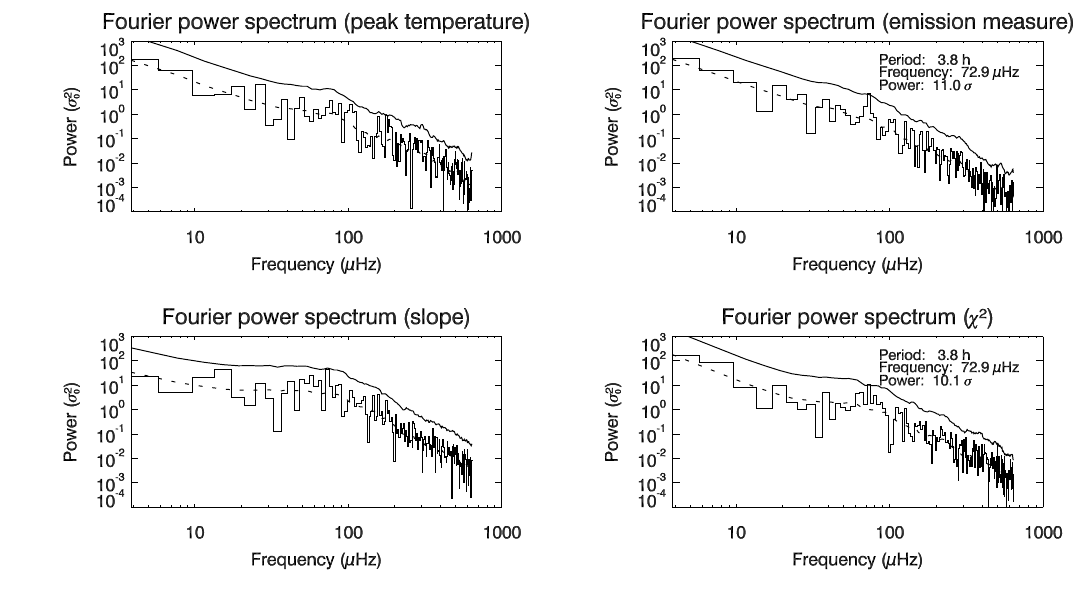}
                 \caption{Fourier power spectra for the three parameters of the DEM model for event 3, same as Fig.~\ref{fig:dem_event_1_spectrum}}
                 \label{fig:dem_event_3_spectrum}
	\end{figure*}

\begin{figure}
		\centering
                 \includegraphics[width=0.5\textwidth]{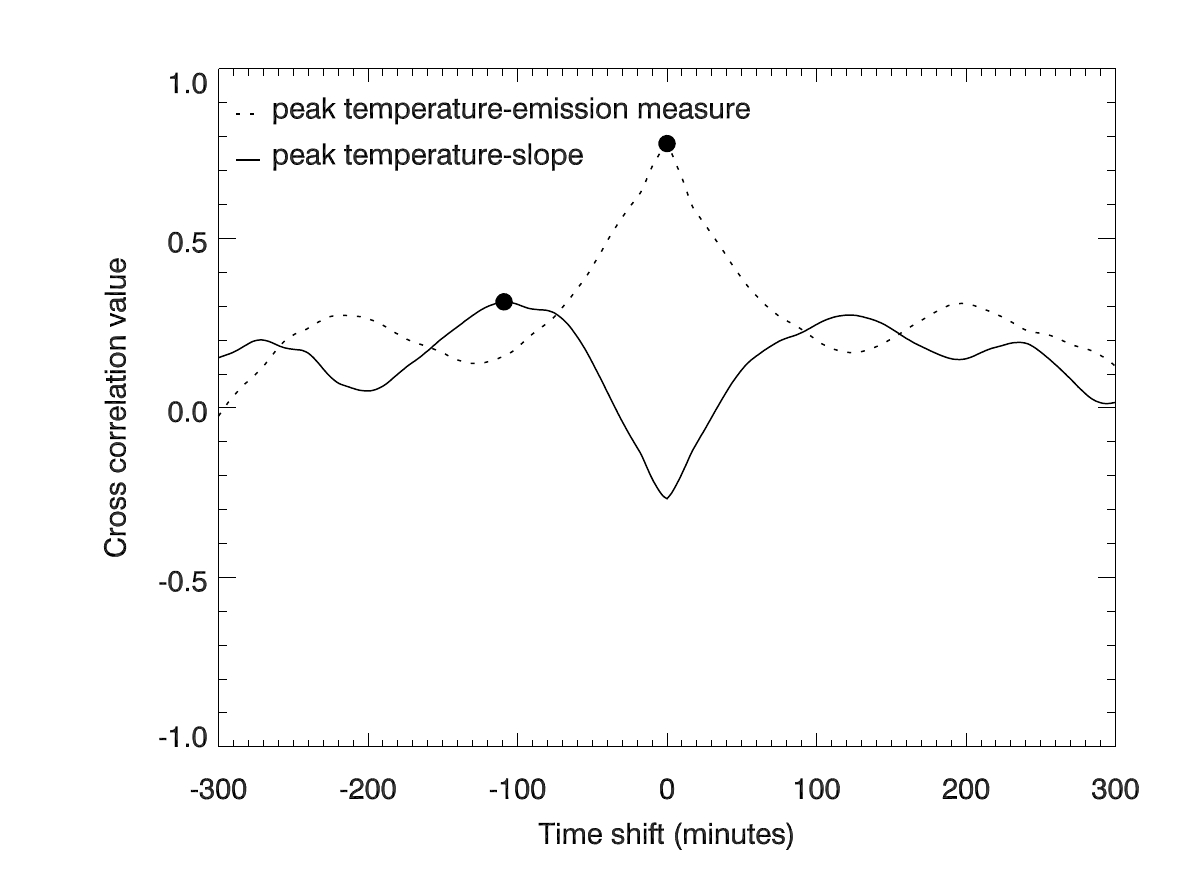}
                 \caption{Same as Fig.~\ref{fig:correlations_dem_event_1}. We used the averaged curves at 1 min of cadence averaged over the small black contour presented in Fig.~\ref{fig:event_1_images} between August 09, 2011 00:00~UT (40 h after the beginning of the long sequence) and August 10, 2011 06:00~UT (70 h after the beginning of the long sequence).}
                 \label{fig:correlations_dem_event_3}
	\end{figure}

Fig.~\ref{fig:dem_event_3_spectrum} represents the power spectra of these parameters. Pulsations are detected for the emission mesure and \(\chi^{2}\) at 72.9~\(\mu\)Hz, the same frequency of detected pulsations in the light curves. The normalized power at these peaks are respectively 11.0\(\sigma\) and 10.1\(\sigma\). If we look under the detection threshold, there is a small peak of power for the the slope (8.4\(\sigma\)). Between 40 hours and 70 hours after the beginning of the large sequence we found a clear correlation (0.78 of peak cross-correlation value) between the peak temperature and the emission measure as seen in Fig.~\ref{fig:correlations_dem_event_3}.

The amplitudes of variations of the DEM parameters are larger than for event~1 and event~2: there are variations from 1.9 to 3.8~MK (a relative amplitude of 99\%) for the peak temperature, from \(3.0 \times 10^{27}\) to \(8.9 \times 10^{27}\)~cm\(^{-5}\) for the emission measure (a relative amplitude of 196\%), and from 1.4 to 3.3 for the slope (a relative amplitude of 144\%). These strong variations of both the peak temperature and the total emission measure are observed in the core of an active region where any type of activity is in general expected to be the highest.

     \subsection{Evidence for widespread cooling}     

\begin{figure*}
		\centering
                 \includegraphics[width=\linewidth, trim = 0 0 0 2.05cm, clip]{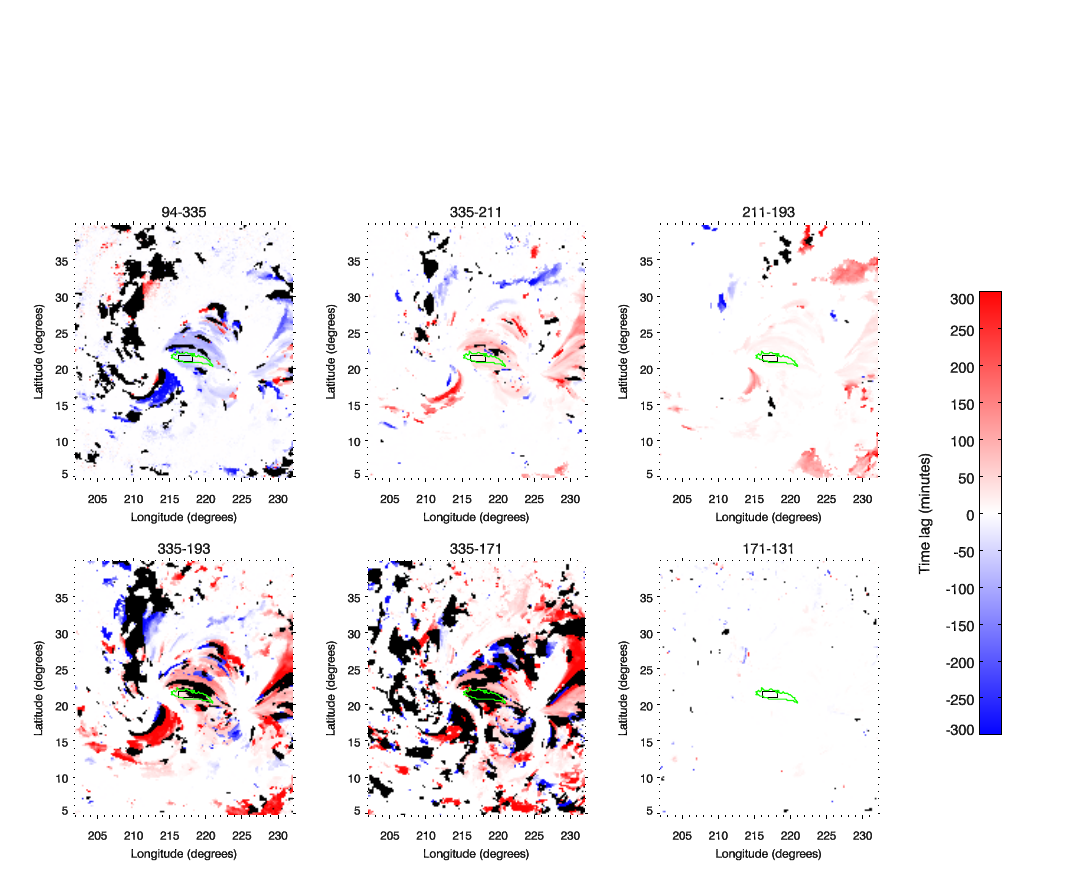}
                 \caption{Same as Fig.~\ref{fig:event_1_lag_maps} for event~3 with the same scale.}
                 \label{fig:event_3_lag_maps}
	\end{figure*}

\begin{figure*}
		\centering
                 \includegraphics[width=\linewidth, trim = 0 0 0 2.05cm, clip]{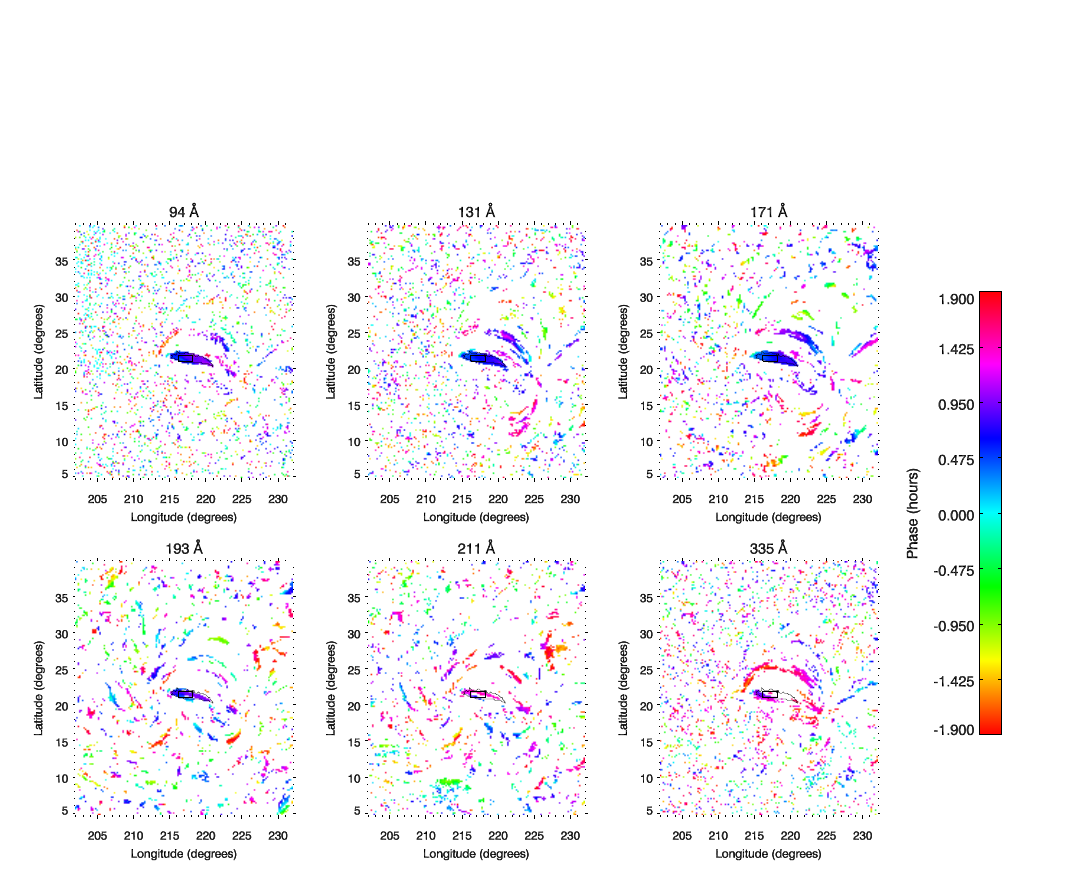}
                 \caption{Same as Fig.~\ref{fig:event_1_phase_maps} for event~3 with the same scale.}
                 \label{fig:event_3_phase_maps}
	\end{figure*}

\begin{figure*}
		\centering
                 \includegraphics[width=\linewidth, trim = 0 0 0 2.05cm, clip]{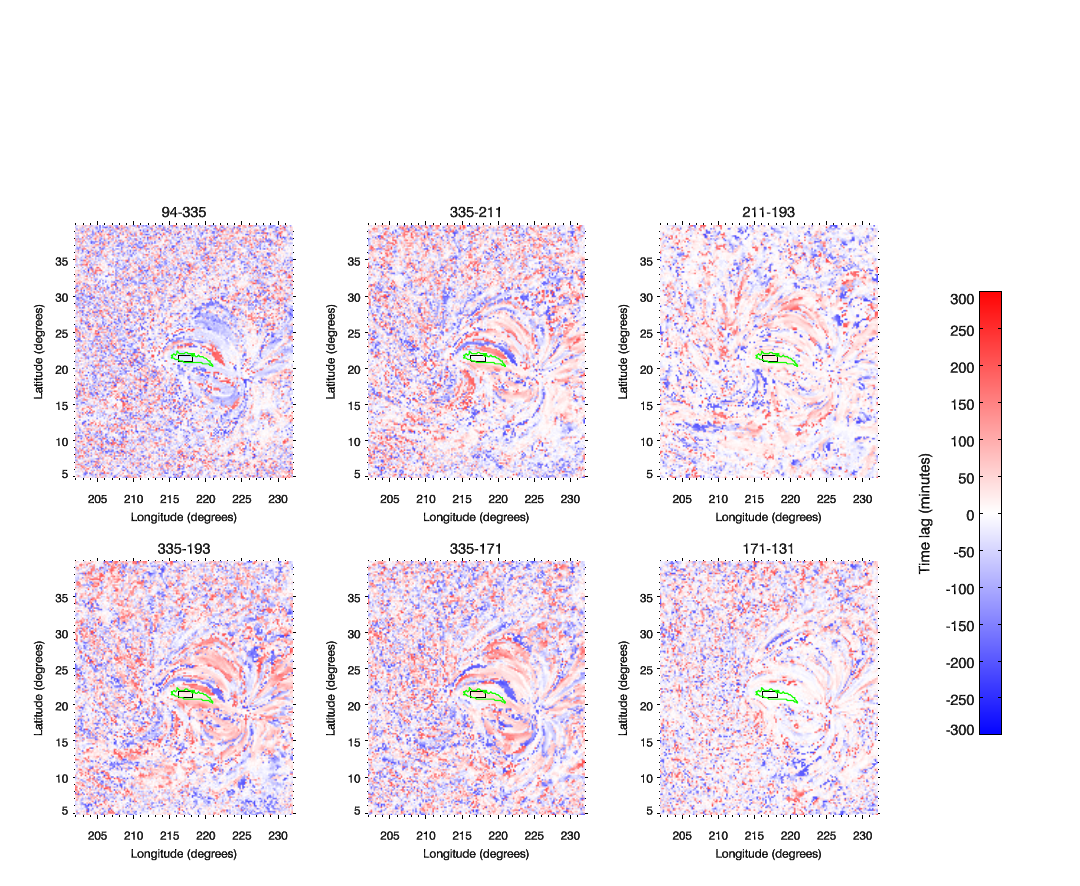}
                 \caption{Same as Fig.~\ref{fig:event_1_phase_maps_lag} for event~3 with the same scale.}
                 \label{fig:event_3_phase_maps_lag}
	\end{figure*}

\paragraph{}The time lags maps in Fig.~\ref{fig:event_3_lag_maps} (made with the peak cross-correlation method) highlight a widespread cooling in this active region. Fig.~\ref{fig:event_3_phase_maps} represents the phase maps. The phases are constant along the loops. Maps produced with the second method (phase differences), shown in Fig.~\ref{fig:event_3_phase_maps_lag}, lead to the same conclusion of widespread cooling. The delay presented in Table~\ref{table:event_3_time_lag_contour} for both methods, show a better succession of channels with the second method: with the first method we have the 335, 211, 94, 193, 171, and 131 order, whereas with the second we obtain 335, 211,193 and 94, 171 and 131, which is the ordering expected for the AIA channels. 
However for this event and for both methods, results are noisy for some pairs of channels. Indeed, for the first method, the peak cross-correlation values are poor for some pairs in the black contour (Fig.~\ref{fig:event_3_lag_maps}) and the time lags for 335-193 and 335-171 are the most affected. The second method suffers from the lack of strong power in the 335, 211 and 193 channels. For event~3, we observe time lags smaller than for event~1 and event~2, that can be explained by the probably smaller loop length (smaller detection area) for event~3.

	\begin{table*}
                \centering
                \caption{Same as Table~\ref{table:event_1_time_lag_contour} for event~3.}
                \label{table:event_3_time_lag_contour}
                        \begin{tabular}{c c c}
                                \hline\hline
                                Pairs of channels &  Time lag (min) &  Time lag (min) \\
                                \, & from cross-correlations (Fig.~\ref{fig:event_3_lag_maps}) & from differences of phase Fig~.(\ref{fig:event_3_phase_maps_lag})  \\
                                \hline

                                335-211 & 26 & 4 \\
                                211-193 & 14 & 29 \\
                                335-193 & 50 & 33 \\
                                94-335 & -36 & -33 \\
                                335-171 & 76 & 42 \\
                                171-131 & 0 & -1 \\
                                
                                \hline
                        \end{tabular}
	\end{table*}

\bibliographystyle{apj}                       
\bibliography{biblio}

\end{document}